\documentclass[a4paper,12pt,leqno]{article}
\makeatletter
\usepackage{amsfonts,delarray,amssymb}
\usepackage[intlimits]{amsmath}
\usepackage{a4,a4wide}

\newtheorem{lem}{Lemma}[section]
\newtheorem{pro}{Proposition}[section]
\newtheorem{defi}{Definition}
\newtheorem{pretheo}{Theorem}

\def\Box{\bullet}

\def\curl{\hbox{curl }}
\def\div{\mathrm{div}  }

\def\noi{\noindent}
\def\La{\Lambda}
\def\be{\begin{equation}}
\def\ee{\end{equation}}
\def\beq{\begin{equation}}
\def\eeq{\end{equation}}

\def\({\left(}
\def\){\right)}
\def\l|{\left|}
\def\r|{\right|}
\def\ep{\varepsilon}
\def\mr{\mathbb{R}}

\def\mc{\mathbb{C}}

\def\lep{|\mathrm{log }\ \ep|}
\def\llep{\log\ \lep}
\def\nab{\nabla}

\def\om{\Omega}
\def\io{\int_{\Omega}}
\def\bo{\partial \Omega}\def\D{\displaystyle}
\def\hal{\frac{1}{2}}
\def\ro{\rho}
\def\he{h_{\rm ex}}
\def\np{\nab^{\perp}}

\def\p{\partial}
\def\hci{{H_{c_1}}}

\def\hcii{{H_{c_2}}}
\def\curl{{\rm curl\,}}
\def\re{{\rho_\ep}}
\def\pe{{\varphi_\ep}}
\def\mue{{\mu_\ep}}
\def\mie{{\mu_\ep^i}}
\def\mje{{\mu_\ep^j}}

\def\re{{\rho_\ep}}

\def\Ae{\boldsymbol{\mathcal{A}_\ep}}
\def\ol{\overline}
\def\pie{{p_\ep^i}}
\def\pje{{p_\ep^j}}

\def\K{\mathcal{K}}
\def\P{\mathcal{P}}
\def\I{\mathcal{I}}
\def\J{\mathcal{J}}
\def\d{\delta}
\def\loc{\text{loc}}
\def\M{\mathcal{M}}
\def\A{\boldsymbol{\mathcal{A}_0}}
\def\Aep{\boldsymbol{\mathcal{A}_\varepsilon}}
\def\Aepp{\boldsymbol{\mathcal{A}_{\varepsilon '}}}
\def\Id{\boldsymbol{\mathcal{I}}}

\def\hb{\overline{h_\ep}}

\numberwithin{equation}{section}

\begin{document}
\title{\huge Pinning phenomena in the Ginzburg-Landau Model of Superconductivity}

\author{{\Large Amandine Aftalion}\\
{\small\it Laboratoire d'Analyse Numérique, B.C.187}\\
{\small \it Universit\'e Pierre et Marie Curie,}\\
{\small \it 4 Place Jussieu,}\\
{\small \it 75252 Paris cedex 05, France.}\\
{\small email: Amandine.Aftalion@ens.fr}
\and
{\Large Etienne Sandier} \\
{\small \it Universit\'e Fran\c cois Rabelais,}\\
{\small\em Département de Mathématiques,}\\
{\small\it Parc Grandmont,}\\
{\small\it 37200 Tours, France.}\\
{\small e-mail: sandier@univ-tours.fr}\\
\and
{\Large Sylvia Serfaty} \\
{\small\it CMLA},\\
{\small\it École Normale Supérieure de Cachan,}\\
{\small\it 61 avenue du Président Wilson,}\\
{\small\it 94235 Cachan Cedex, France.}\\
{\small e-mail: serfaty@cmla.ens-cachan.fr}}

\maketitle
\begin{abstract}
We study the Ginzburg-Landau energy of superconductors 
with a term $a_\ep$ modelling the pinning of vortices by impurities 
in the limit of a large Ginzburg-Landau parameter $\kappa=1/\ep$.
The function $a_\ep$ is oscillating between 1/2 and 1 with a scale
which may tend to 0 as $\kappa$ tends to infinity.

Our aim is to understand that in the large $\kappa$ limit, 
stable configurations should correspond to vortices pinned at 
the minimum of $a_\ep$ and to derive the limiting homogenized 
free-boundary problem which arises for the magnetic field in 
replacement of the London equation. 
 The method and techniques that we  use are inspired from those of  \cite{ss3} (in which the case $a_\ep \equiv 1$ was treated) and based on energy estimates, convergence of measures and construction of approximate solutions. Because of the term $a_\ep(x)$ in the equations, we also need homogenization theory to describe the fact that the impurities, hence the vortices, form a homogenized medium in the material.

\end{abstract}


\section{Introduction}
Superconducting materials have the property of expelling an applied magnetic field. In fact, the behaviour of a superconducting sample varies according to the value of the applied field and the value of the Ginzburg-Landau parameter $\kappa$ which is characteristic of the material. When $\kappa$ is large, the superconductors are known as type-II and display vortex patterns for intermediate fields: for high magnetic fields, the material is normal and the magnetic field penetrates into the sample, for  low fields, the material is superconducting, that is the magnetic field is expelled from the sample and for intermediate fields, there are vortices.
 The vortex state is a state where
 the superconducting and the normal phases coexist: at the center of the vortex, the material is normal and the vortex is circled by a superconducting current carrying  a quantized amount of magnetic flux. 
 The motion of vortices generates an electric field hence energy-dissipation. In order to have the desired property of dissipation-free current flow, the vortices have to be held fixed or pinned. In practice, attempts are made to pin vortices either by varying the thickness of the material or by introducing impurities or normal inclusions.  Sufficiently strong pinning is necessary for functional superconductors capable of sustaining strong currents and high magnetic fields. The new high-temperature (high $T_c$) superconductors are strongly type-II superconductors, that is their phenomenology is dominated by the presence and properties of vortices when an exterior magnetic field is applied. The pinning problem is particularly intricate in high-$T_c$ superconductors where it depends on specific structures such as layering and structural defects. 

In this paper, we will be concerned with the case where the vortices are pinned by impurities in the framework of the Ginzburg-Landau model. We will study the behaviour of global minimizers of  the Ginzburg-Landau energy when a term modelling the pinning of vortices by impurities is added, in the limit of a large Ginzburg-Landau parameter $\kappa$, which describes extreme type-II materials.

\subsection{The Ginzburg-Landau model with a pinning term}

Recall that in the framework of the Ginzburg-Landau theory (see \cite{t} for more details), the state of the material is completely described by a vector potential $A$ and a complex-valued function $u$, which can be thought of as a wave-function of the superconducting electrons, and is nondimensionalized such that $|u|\leq 1$.    The type of material is characterized by the Ginzburg-Landau parameter $\kappa$ and in the case of type II, $\kappa$ is large so that we define $\ep=1/\kappa$, which will be small. 
The energy is the following:
\begin{equation}\label{GL}
J_\ep(u,A) = \frac{1}{2} \int_{\om} |(\nabla -i A) u|^2  +\frac{1}{2\ep^2}\(a_\ep(x)-|u|^2\)^2 + |h-\he|^2.
\end{equation}
Here, $\om$ is the domain occupied by the superconductor, $h=\curl A$ is the magnetic field and $ \he$ is the exterior magnetic field which is constant in our problem.
 A common  simplification is to restrict to a two-dimensional problem corresponding to an infinite cylindrical domain of section $\om\subset \mr^2$ (smooth and simply connected), for an applied field parallel to the axis of the cylinder.
 Then $A:\om \mapsto \mr^2 $, $h$ is real-valued and all the quantities are translation-invariant.

 The energy $J_\ep$ that we are going to study here is slightly different from the classical Ginzburg-Landau energy in the sense that there is a term penalizing the variations of the order parameter $u$. We denote this function by $a_\ep(x)$. In the case originally studied by Ginzburg and Landau, $a_\ep \equiv 1$. In this paper, a typical example for $a_\ep$ would be to oscillate between $1/2$ and 1 in the domain, with a typical scale $\eta$ which may tend to 0 with $\ep$. The minima of $a_\ep$ correspond to the impurities in the material. Hence it is expected that these minima will be the pinning sites for the vortices.

 The modified Ginzburg-Landau functional (\ref{GL}) was first written down by Likharev \cite{l}.  Then, this model has been used and developed in  \cite{cr} and \cite{cdg}. Review articles on the topic include \cite{bfglv}, \cite{c1}, \cite{c2} and \cite{p}.
Computational evidence that the vortices are attracted by the impurities, that is the points of minimum of $a_\ep(x)$ can be found in \cite{cdg} or \cite{dgp}.

In this paper, we want to address the question of how the term $a_\ep$ will modify the properties of the superconductor in the presence of an exterior magnetic field.
 The method and techniques that we are going to use are inspired from those of  \cite{ss3} (in which the case $a_\ep \equiv 1$ was treated) and based on energy estimates, convergence of measures and construction of approximate solutions. Because of the term $a_\ep(x)$ in the equations, which can be a rapidly oscillating function, we will also need homogenization theory (\cite{cd}, \cite{jo}, \cite{mut}) to describe the fact that the impurities, hence the vortices, form a homogenized medium in the material.

\subsection{The equation for the magnetic field}
The Ginzburg-Landau equations associated to the functional (\ref{GL}) when minimizing for $\{(u,A)\in H^1(\om, \mc)\times H^1(\om, \mr^2)\}$ are
$$
\leqno(\mathrm{G.L.} )
\left\{ \begin{array}{l} -(\nab-iA) ^2 u  = \displaystyle{\frac{1}{\ep^2}}u(a_\ep(x)-|u|^2) \\
             -\np h=<iu,(\nab -iA) u>, \end{array} \right.
$$
with the boundary conditions
$$ \left\{\begin{array}{ll}  h=\he & \text{on} \ \bo\\
(\nab u - i A u) \cdot n =  0  & \text{on} \ \bo.\end{array} \right.$$
Here $\np $ denotes $(-\p_{x_2}, \p_{x_1})$, and $<z,w>=Re (z\overline{w})$ for $z,w$ in $\mc$. Recall that the problem is invariant under the gauge transformations
$$
\left\{ \begin{array}{l}
u\to ue^{i\Phi}\\
A\to A+\nab \Phi,
\end{array}
\right.$$
where $\Phi \in H^2(\om, \mr)$.
Physically meaningful quantities are  gauge invariant. These include the energy $J_\ep$, the magnetic field $h$ and the superconducting current $j=<iu,(\nab -iA) u>$.

\hfill

Let us describe the properties of a superconductor. These phenomena are described for instance in \cite{t}.
The state of the material depends on the applied field $\he$. In the absence of pinning, that is when $a_\ep \equiv 1$, there are two critical fields $\hci$ and $\hcii$ for which a phase transition occurs. Above $\hcii=O(\frac{1}{\ep^2})$, superconductivity is destroyed and the material is in the normal phase $(u\equiv 0, h\equiv \he)$. Below $\hci= O(\lep)$, the material is superconducting everywhere, that is $|u|\thicksim 1$. This is the Meissner phase characterized by complete expulsion of the magnetic field : in the limit when $\ep$ goes to zero, the magnetic field satisfies the London equation
\beq\label{london}
\left\{ \begin{array}{ll}
-\Delta h + h = 0 & \text{in} \ \om\\
h= \he & \mathrm{on} \ \bo.\end{array} \right.\end{equation}
Between $\hci$ and $\hcii$, the material is in the mixed phase defined by the coexistence of the normal and superconducting phases in the form of vortex filaments: the magnetic field penetrates into the material in the form of flux lines at the center of which $u$ vanishes.
The induced magnetic field approximately satisfies
\beq\label{londond}
\left\{ \begin{array}{ll}
-\Delta h + h = 2\pi \sum_i d_i \delta_{p_i} & \text{in} \ \om\\
h= \he & \mathrm{on} \ \bo,\end{array} \right.
\end{equation}
where the $p_i$'s are the centers of the vortices, and the $d_i$'s their degrees, that is the topological degree of the map $u/|u|$. These filaments are of characteristic size $\ep$. They are surrounded by a  superconducting region in which $|u|\thicksim 1$. In order to minimize their repulsion, the flux lines form a triangular lattice, called the ``Abrikosov lattice''. With increasing fields, the density of flux lines increase until the vortices overlap and $\hcii$ is reached.  The generation of vortices by the external field  has been mathematically studied very recently in \cite{s1,s2,s3,ss1,ss2,ss3}.

In \cite{ss3}, it is proved among other things that, in the limit when $\ep$ tends to 0, equation (\ref{londond}) is replaced by
 \beq\label{hss3}
- \Delta h_*+h_*=\mu_*
\end{equation}
where $\mu_*$ is the density of vortices in units of $\he$ and $h_*=h/\he$. The measure $\mu_*$ is supported in an inner region $\omega$ depending on the value of $\he$ and is of uniform density in $\omega$.

\hfill

Our aim is to give a rigorous proof that in the small $\ep$ limit, stable configurations should correspond to vortices pinned at the minimum of $a_\ep$ and to derive the limiting homogenized free-boundary problem which arises for the magnetic field in replacement of the London equation (\ref{hss3}).

\hfill

Using the second equation in (G.L.), we notice that the energy can be rewritten
\beq\label{en2}
J_\ep (u,A)= \frac{1}{2} \int_{\om} \frac{1}{|u|^2} |\nab h|^2 +|h-\he |^2+\frac{1}{2} \int_{\om}|\nab |u||^2+
\frac{1}{2\ep^2}(a_\ep(x)-|u|^2)^2 .
\end{equation}
We will show that for a sequence of minimizers $(u_\ep,A_\ep)$, the second integral in (\ref{en2}) is negligible.
Then, when $\ep$ tends to 0, $|u|^2\thicksim a_\ep(x)$ outside the vortices, and our main result will state that $h_\ep=\curl A_\ep$ satisfies roughly the following equivalent of (\ref{londond}) in the case of pinning:
\beq\label{hep}
- \div \left(\frac{1}{a_\ep} \nab h_\ep\right)+h_\ep=2\pi \sum_i d_i \delta_{p_i}.
\end{equation}
The existence of pinning will modify the locations $p_i$ of the vortices and the value of $\hci$. 

Since $a_\ep$ is a rapidly oscillating function describing  impurities, the framework for passing to the limit when $\ep$ is small is that of homogenization theory.
 When passing to the limit in (\ref{hep}), we obtain a different limiting operator from (\ref{hss3}), that is
\beq\label{lp}
- \div \( \A \nab h_*\) +h_*=\mu_*
\end{equation}
where $\mu_*$ is a positive measure which is supported in an inner domain $\omega_\Lambda$ and $\A$ is the homogenized limit of the matrix $\Aep =\displaystyle{\frac{1}{a_\ep} \Id}$ in the sense of $H$-convergence, see definition below.

\begin{defi}\label{Hconv}
We say that the family  of $2\times 2$ matrices $\Aep$ $H$-converges to $\A$ when $\ep $ tends to 0, if and only if, for any $f$ in $H^{-1}(\Omega)$, the solution $v_\ep$ in $H^1_0(\om)$ of
$$-\div(\Aep \nabla v_\ep)+v_\ep=f$$ satisfies
$$\begin{array}{ll}
v_\ep \mathop{\rightharpoonup}v_0 \quad\hbox{weakly in}\quad H^1_0(\Omega),\\
\Aep \nabla v_\ep\mathop{\rightharpoonup}{\A}  \nabla v_0\quad\hbox{weakly in}\quad\( L^2(\Omega)\)^2,
\end{array}$$
where $v_0$ is the $H^1_0(\om)$ solution of $$-\div({\A}  \nabla v_0)+v_0=f.$$
\end{defi}
We refer to the work of 
 Murat and Tartar \cite{mut} for more details on the notion of $H$-convergence; one can  also see \cite{cd,jo}. In the following, we will always let ${\Aep} =\D\frac{1}{a_\ep} \Id$. 
 Then ${\A}$ is also a diagonal matrix. In the general case, the computation of $\A$ is hard and not always known, see \cite{jo} for  examples. But in some simple cases, this definition allows to compute $\A$. 
 For instance, if $a_\ep(x)=a(x/\ep)$, and $a(x)=a_1(x_1)a_2(x_2)$ where $a_1$ and $a_2$ are periodic, then
$$\A=diag\Bigl( \frac{1}{a_1^0},\frac{1}{a_2^0}\Bigr), \quad \hbox{with}\quad a_i^0=\overline{a_i}\overline{\Bigl( \frac{1}{a_j}\Bigr)}$$
where $\overline{a_i}$ denotes the  mean of $a_i$ over a period (see \cite{jo}).
Note that even though the sequence $a_\ep$ has no pointwise limit, the limiting problem and $\A$ are well defined.

An important property of $H$-convergence (see \cite{mut}) is that if
  the sequence $a_\ep$ is bounded from below and above by positive constants independent of $\ep$, then there exists a subsequence ${\Aepp}$ and a matrix $\A$ for which ${\Aepp} $ $H$-converges to $\A$. For us, it will imply in the following that up to the extraction of a subsequence, the family $\Aep$ $H$-converges to some limit $\A$, thus leading to  the limiting problem (\ref{lp}).

\subsection{Main results}

Let us now state our hypotheses and results.
 We assume that $\he$ is a function of $\ep$ and that the following limit exists and is finite:
\beq\label{La}
\La= \lim_{\ep \to 0 } \frac{\lep}{\he(\ep)}.
\end{equation}
Moreover, we make the following hypotheses on the function $a_\ep(x)$:
\begin{itemize}
\item[(H1)] There exists a constant $b_0>0$ such that $b_0\leq a_\ep(x)\leq 1$.
\item[(H2)] There exist a constant $C$ and a sequence $\eta(\ep)$ (which may tend to 0 with $\ep$) such that $1/\eta(\ep) \ll \he$ and
$|\nabla a_\ep |\leq \D\frac{C} {\eta(\ep)}$.
\item[(H3)] There exist a continuous function $b(x)$ 
 and  a nonnegative functions $\beta_\ep(x)$ such that $a_\ep(x)=b(x)+\beta_\ep(x)$ and for any $\ep>0$ and any $x\in\om$, $\min_{B(x,\d(\ep))}\beta_\ep =0$, where 
$$ \d(\ep) \ll \frac{1}{(\llep )^\hal}.$$
\item[(H4)] The family of matrices $\Aep$ $H$-converges to ${\A}$.
\end{itemize}
Note that, as we mentioned earlier, it follows from hypothesis (H1) and the compactness of the set of matrices bounded from above and below that there exists a subsequence of $\Aep$ which $H$-converges to ${\A}$ \cite{mut}. Our hypothesis (H4) is there to restrict to this subsequence for ease of notation and to impose that the whole sequence converges.
 Moreover, (H2) means that $a_\ep$ can be a constant independent of $\ep$ but can also oscillate very quickly with $\ep$ (but not too quickly, i.e. not quicker than $\he$).
 Note that in the case where $a_\ep$ does not depend on $\ep$, then ${\Aep}=\A$ is constant.

Let us emphasize that because $\beta_\ep\geq 0$, $b$ can be thought of as the lower envelope of $a_\ep$ and the local minima of $a_\ep$ are the local  minima of $b$. Hence $b$ will be related to the pinning sites of vortices and the oscillations of $a_\ep$ are those of $\beta_\ep$. Moreover, the hypotheses imply that $b\geq b_0$.

\hfill

First, let us state the result concerning the limiting problem (\ref{lp}). We relate $h_*$ and $\mu_*$ to the minimum of a variational problem. Let ${\cal M}$ denote the space of Radon measures in $\Omega$.

\begin{pretheo}\label{theo1}
Let us assume that (H1) to (H4) are satisfied. Let us define for any $\La \geq 0$,
\begin{equation}\label{E(f)}
E(f)= \frac{\Lambda}{2}\io b(x)\l|-\div  ({\A}\nabla f) + f\r|+ \hal\io \nabla f \cdot {\A}\nabla f + \l|f-1\r|^2,
\end{equation}
over $$V=\{ f \hbox{ s.t. } f-1\in H^1_0(\om), \hbox{ and } -\div  ({\A}\nabla f)+ f\in {\cal M}\}.$$ 
The minimizer $h_*$ of $E$ over $V$ exists and is unique.  It satisfies
$$\leqno{(\mathrm{P})} \left\{ \begin{array} {l}
h_*-1 \in H^1_0(\om)\\
\mu_*= -\div  ({\A}\nabla h_*) + h_*\in {\cal M}\\
h_* \ge 1-\D\frac{\Lambda b}{2} \ \mathrm{in} \ \om\\
 
\mu_* \left( h_*-(1-\D\frac{\Lambda b}{2})\right)=0 \ \mathrm{in} \ \om .\end{array} \right. 
$$
Moreover $\mu_* \geq 0$ and $\mu_*\in H^{-1}(\Omega)$.
\end{pretheo} 
Problem (P) is a free-boundary problem, called in the literature an ``obstacle problem'' (see \cite{r}).
Another way of considering problem (P) is to define the subset of $\om$
\beq\label{omega}
\omega_\Lambda=\{x \in \om,\hbox{ s.t. } h_*=1-\La b/2\}.\end{equation}
Then $\mu_*=0$ in $\Omega\setminus\overline{\omega_\Lambda}$, and $h_*=1-\La b/2$ in $\omega_\Lambda$, $\p \omega_\Lambda$ being called the ``free-boundary'', because $\omega_\Lambda$ is unknown and uniquely determined by the set of equations (P).

Note that if ${\A}$ and $b$ are smooth enough then $h_*$ is $C^{1,\alpha}$ ($\alpha<1$), $\mu_*$ is in $L^\infty$,  the free-boundary $\p \omega_\Lambda$ is regular for almost every $\Lambda$ (see \cite{bm}) and then we can write
$$\mu_*=1-\D\frac{\Lambda b}{2}+\D\frac{\Lambda}{2}\div({\A} \nabla b)\quad \hbox{in}\quad \omega_\Lambda.$$
Once we have proved Theorem \ref{theo1} concerning the limiting problem, we can get convergence for any sequence of minimizers $(u_\ep,A_\ep)$ of the energy $J_\ep (u_\ep,A_\ep)$ to $E(h_*)$ in a sense similar to $\Gamma$-convergence.

\begin{pretheo}\label{theoconv}
Let us assume that (\ref{La}) and (H1) to (H4) are satisfied. Let $(u_\ep,A_\ep)$ be a family of minimizers of $J_\ep$, and $h_{\ep} = \curl A_{\ep}$  the associated  magnetic field.  Then, as $\ep$ tends to 0,
$$\frac{h_{\ep}}{\he}\to h_* \quad \text{weakly in} \ H^1(\om),$$ 
where $h_*$ is  the minimizer of $E$. Moreover,
\begin{eqnarray}
&&\lim_{\ep \to 0}\frac{J_\ep(u_{\ep}, A_{\ep})}{\he^2} = E(h_*)=\frac{\Lambda}{2}\io b|\mu_*|+\hal \io \nab h_*\cdot {\A}\nab h_*+|h_*-1|^2,\\
&&\frac{|\nabla h_{\ep}|^2}{\he^2 a_\ep} \to \nab h_*\cdot {\A}\nab h_*+  \Lambda b \mu_*,
\quad\hbox{in the sense of measures.}
\end{eqnarray}
\end{pretheo}
One can easily notice that if $\Lambda= 0$ (i.e. if $\he \gg\lep$), the solution of (P) is $h_* = 1$, and $E(h_*)=0$. In this case,  Theorem \ref{theoconv} asserts that 
$$\frac{h_\ep}{\he} \to  1\quad  \text{strongly in}\ H^1,\quad \text{and}\quad \lim_{\ep \to 0}\frac{\min J_\ep}{\he^2}= 0.$$ 
The proof of Theorem 2 is the main part of the paper (see Section I.6 for a sketch).

\subsection{The case $\Lambda>0$}

Let us now present some stronger results in the case where $\La$ is positive, i.e. $\he$ is of the order of $\lep$.
 The first issue is to determine mathematically the location of vortices. From the physics, we know that vortices are the zeroes of $u_\ep$ with non-zero winding number. Instead of defining vortices, we isolate them in disjoint vortex balls covering the set where $|u_\ep|$ is small. The centers of these balls can be thought of as being the centers of the vortices.

 \begin{pro} \label{propballs}
Let us assume that $\Lambda>0$ and that (H1) to (H4) are satisfied, then
 there exists $\ep_0 $ such that if $\ep<\ep_0$ and $(u_\ep,A_\ep)$ is a minimizer of $J_\ep$,  there exists a family of balls of disjoint closures (depending on $\ep$) $(B_i)_{i \in I_{\ep}} = (B(p_i, r_i))_{i \in I_{\ep} }$ satisfying~:
\begin{eqnarray}\label{usmall}
&&\left\{ x\in \om, \, |\sqrt{ a_\ep(x)} - |u_{\ep}(x)|| \ge \frac{1}{\lep}\right\} \subset \bigcup_{i \in I_{\ep} } B(p_i, r_i). \\\label{sumri}
&& \sum_{i \in I_{\ep} } r_i \le \frac{1}{e^{\sqrt{\lep}}}\\\label{intbi}
&&\hal\int_{B_i} \D\frac{|\nabla h_\ep |^2}{|u|^2} \ge \pi b(p_i)|d_i| \lep (1 - o(1)),\end{eqnarray}
where $h_\ep = \curl A_\ep$, and $d_i = deg( \frac{u_{\ep}}{|u_{\ep}|} , \p B_i)$ if $\overline{B_i}\subset \om$, and 0 otherwise.
\end{pro}
This proposition will be proved at the beginning of Section II. Here is the meaning of the different inequalities: (\ref{usmall}) locates the set where $|u_\ep|$ differs from $a_\ep$, which is contained in a union of disjoint balls; these balls represent the vortices or clusters of vortices. (\ref{sumri}) gives a control on the size of the balls and (\ref{intbi}) gives a lower bound on the energy, which is the contribution of vortices according to their degree $d_i$ and their location $p_i$, appearing through the value $b(p_i)$. As opposed to the case of $a_\ep \equiv 1$ (see \cite{ss3}), the least energy is attained for $p_i$ at the minimum of $b$.

Using this proposition, Theorem \ref{theo1} can be made more precise:
\begin{pretheo}
Let us assume that $\Lambda>0$ and that (H1) to (H4) are satisfied.
For any balls $B(p_i,r_i)$ and integers $d_i$ which satisfy (\ref{usmall})-(\ref{sumri})-(\ref{intbi}), then 
\begin{eqnarray}\label{convaepi}
\lim_{\ep\to 0}\frac{2\pi}{\he}\sum_{i\in I_{\ep}} d_i a_\ep({p_i})  &= \quad\io b |\mu_*|,\\
\label{ddelta}
\frac{2\pi}{\he}\sum_{i\in I_{\ep}} d_i \delta_{p_i}  &\D\mathop{\longrightarrow}_{\ep \to 0}\quad\mu_*,\\
\label{|d|delta}
 \frac{2\pi}{\he}\sum_{i\in I_{\ep}} |d_i| \delta_{p_i} & \D \mathop{\longrightarrow}_{\ep \to 0}\quad \mu_*,\end{eqnarray}
in the sense of measures,  where $$\mu_*= -\div ({\A} \nabla h_*)+h_*  .$$ 
\end{pretheo}

\subsection{Physical interpretations and consequences}

Our results show that $h_*\he$ is a good approximation of $h_\ep$ and that, in the limit $\ep \to 0$, the vortices are scattered in an inner region $\omega_\Lambda$ with density $\mu_*$, where $h_*=1-\Lambda b(x) /2$. In the outer region $\Omega \setminus \overline{\omega_\Lambda}$, there are no vortices and $h_*$ satisfies $-\div(\A \nab h_*)+h_*=0$. Unlike the case $a_\ep \equiv 1$, the vortex-density in $\overline{\omega_\Lambda}$ is non-uniform in general.  Moreover, as $\Lambda$ decreases, the  vortex-region first appears at the minimum of $\psi$ as defined by problem (\ref{psi}) below: as in \cite{ss3}, we can derive a necessary and sufficient condition for $\omega_\Lambda$ to be nonempty.
\begin{pro}\label{prophci}
Let $\psi$ be the solution of
\begin{equation}
\label{psi}
\left\{  \begin{aligned}
         -\div(\A\nabla \psi) + \psi &= -1 & \quad & \mathrm{in}\ \om \\
          \psi &=0 &\quad & \mathrm{on} \ \bo,
\end{aligned}\right.  
\end{equation}
then
$$\omega_\Lambda \neq \varnothing \Longleftrightarrow  \lim_{\ep\to 0}\frac{\he }{|\log \ep|} \geq \frac{1}{2 \max |\psi|}.$$
\end{pro}
If we define  $\hci$ as the field such that
for $\he \leq \hci$, the minimizer of the energy has no vortex (i.e. $|u|\geq b_0 /2$)
and  for $\he \geq \hci$, there exists a minimizer with vortices; 
then Proposition \ref{prophci} gives a hint that 
$$\hci \simeq \frac{|\log \ep|}{2\max |\psi|}.$$
Thus the presence of pinning modifies the values of the first critical field (see \cite{s1,ss1} for the case without pinning).
In fact, we could adjust the proof of \cite{ss1} to obtain: there exists $k_\ep=O(|\log |\log \ep||)$ such that for $\ep$ small enough and 
$$\he \leq \frac{|\log \ep|}{2\max |\psi|} -k_\ep$$
then any minimizer has no vortex. 

Furthermore, the position of the  minimum of $\psi$ depends on the pinning potential $a_\ep(x)$. As   $\Lambda$ further decreases, corresponding to $\he$ increasing, the vortex-region $\omega_\Lambda$ grows, until, for $\Lambda=0$ ($\he \gg\lep$), $\omega_\Lambda=\om$. At this point there are so many vortices that the macroscopic density of vortices and the induced magnetic field are no longer influenced by $a_\ep$. In other words, the strength of flux pinning is 0 for $\he\gg |\log \ep|$.

\hfill 

In the case where $a_\ep (x)=a(x)$ is independent of $\ep$, $a(x)= b(x)$ and $\A=a^{-1}\Id$. Hence the limiting problem is a London equation with weight. We would like to point out that it is natural to define a vortex velocity by
 $v=\frac{1}{|u|^2}\nabla h$ (see \cite{cp}). In particular
$$v_*=\frac{1}{a}\nabla h_*$$
can be defined as a limiting velocity (per unit of $\he$). Note that in $\omega_\Lambda$, since $h_*=1-\hal\Lambda a $, then $v_*=-\hal\Lambda \nabla \log a $. It implies that when $a$
 is constant, $v_*=0$ and there is no mean current in the vortex region. But when $a$ varies spatially, there is a nonzero limiting mean current and a nonzero limiting velocity $v_*$. Hence $v\simeq \he v_*$ that is $\hal\log \kappa \nabla \log a $. This is 
 the result of Chapman-Richardson \cite{cr} in the case where the three-dimensional vortex line has no curvature. They describe the phenomenon saying that the variation in $a$ acts as a pinning potential.

When $\Lambda=0$, the velocity $v_*$ is zero as well. Decreasing $\Lambda$ means increasing the field. So when $a$ varies spatially, there is a critical exterior magnetic field above which the pinning potential has no role and the current is destroyed.

\hfill

In the general case where $a_\ep$ depends on $\ep$, it would be interesting to prove a convergence of the mean vortex velocity $v_\ep=\frac{1}{|u_\ep|^2}\nab h_\ep$. 
Still, one can observe  two different effects coming from the presence of pinning in the term $|\nabla h_\ep|^2 /a_\ep$
 and resulting in the energy $E(h_*)$ in the homogenization process:
\\
-- One effect is related to the concentration of energy in the vortices and  the location of the vortices. It appears through the term
$$\frac{\Lambda}{2}\io b |\mu_*|$$
in the limiting energy $E$. This term is smaller if $\mu_*$ is non-zero at points where $b$ is minimal. (\ref{convaepi}) implies that vortices go to points where $\beta_\ep =0$. These points will be called pinning sites in the following. Because $\d(\ep)$ tends to 0, the number of such points is big.
 The effect on the position of vortices is to see $b$ and the minima of $b$. Moreover,  since (\ref{ddelta}) and (\ref{|d|delta}) have the same limit, it means that vortices tend to  have positive degrees.

If $b$ does not depend on $x$ then $h_*$ and $\mu_*$ are constant in $\omega_\Lambda$, and there is no change for the location of vortices from the case $a_\ep \equiv 1$. On the other hand, if $b$ is non-uniform, then $\nabla h_*$ is non-constant in $\omega_\Lambda$ and there is a pinning current.
  If for example the domain is a disc and the minima of $b$, that is the impurities, are located at sites different from the center of the disc, one expects  that vortices, or the vortex-region $\omega_\Lambda$ will be closer to the minima of $b$, but it seems difficult to give a rigorous proof of this qualitative fact.  

\hfill

\noi
-- The other effect is due to the rapid oscillations of $a_\ep$ with $\ep$ and comes from the energy outside the vortices, converging to the homogenized term 
$$\hal \io \nab h_* \cdot \A \nab h_*+|h_*-1|^2$$ in $E$.
 It changes the equation for the magnetic field $h$ from the usual London equation. If $\beta_\ep \neq 0$, then the homogenization effect can be anisotropic.
 The size $\d(\ep)$ (which can be related to $\eta$ if $\beta_\ep$ is not identically 0) cannot be taken bigger than in (H3), otherwise each pinning site would be too large and the vortices could push one another outside the pinning site.

\hfill

Let us also point out that we cannot allow stronger oscillations of $a_\ep$ than in (H2), because the second integral in (\ref{en2}) would become the dominant term. It would be interesting to investigate what happens if (H2)-(H3) are relaxed.

\subsection{Main steps of the proof}

Let us now state the two steps of the proof of Theorem \ref{theoconv}. It is obtained as in \cite{ss3} by getting first a  lower bound on the energy, Proposition \ref{proplowerbd}, proved in Section II,  and then an upper bound, Proposition \ref{propupperbd}, proved in Section III.

\begin{pro}\label{proplowerbd}
Let us assume that $\La >0$ and that (H1) to (H4) are satisfied. Let $(u_\ep,A_\ep)$ be a minimizer of $J_\ep$. Then
\begin{equation}
\label{eqlowerbd}
\liminf_{\ep\to 0}\frac{1}{\he^2}  J_\ep(u_\ep,A_\ep)\ge \frac{\Lambda}{2}\io b|\mu_*|+\hal \io \nab h_*\cdot {\A}\nab h_*+|h_*-1|^2,
\end{equation}
where $h_*$ is the solution of $(P)$.
\end{pro}
 
\begin{pro}\label{propupperbd}
Let us assume that $\La >0$ and  that (H1) to (H4) are satisfied. Let $\mu$ be a positive Radon measure, and let $(u_\ep,A_\ep)$ be a minimizer of $J_\ep$. Then
\begin{equation}
\label{equpperbd}
\limsup_{\ep\to 0}\frac{1}{\he^2}  J_\ep(u_\ep,A_\ep)\le \frac{\La}{2}\io b \ d\mu +\hal\io \nabla h \cdot \A \nabla h +|h-1|^2,
\end{equation}
where $h$ is the solution of
\beq\left\{\begin{aligned}\label{G0}
- \div( {\A} \nabla h)+h=\mu \quad \hbox{in}\quad \Omega,\\
h =1  \quad \hbox{on}\quad \partial\Omega.
\end{aligned}\right.\end{equation}
\end{pro}

Section II is devoted to the proof of Proposition \ref{proplowerbd}. Let $(u_\ep, A_\ep)$ be a sequence of minimizers and $h_\ep =\curl A_\ep$. The energy $J_\ep (u_\ep,A_\ep)$ gives two contributions: inside the vortex balls and outside. Thus,
 first we prove Proposition \ref{propballs} where the vortex balls $B_i$ with centers $p_i$ are constructed and where the vortex energy is bounded from below. We define 
\begin{equation}
\label{23}
\mu_\ep=\frac{2\pi}{\he} \sum_{i\in I_\ep} d_i \delta_{p_i}.\end{equation}
Then, Proposition \ref{propballs} implies
\begin{equation}
\label{24}
\frac{1}{\he^2} \int_{\cup_{i \in I} B_i} \frac{1}{|u|^2} |\nabla h_\ep|^2\geq \frac{|\log \ep|}{\he}\io b|\mu_\ep|,\end{equation}
which gives the lower bound inside the vortex balls.
The next step is to pass to the limit in the energy outside the vortex balls. Letting $h_0$ be the weak $H^1$ limit of $h_\ep / \he$, we obtain the following, which is similar to a standard result in homogenization theory
\begin{equation}
\label{25}
\liminf_{\ep\to 0} \int_{\om \backslash \cup_i B_i} \frac{|\nab h|^2}{a_\ep\he^2} \ge \io \nab h_0\cdot \A \nab h_0.\end{equation}
This requires to introduce an auxiliary problem before applying the homogenization theory result and it works because the vortex balls are small and thus can be taken out of the first integral.

Finally we derive from the Ginzburg-Landau equations the crucial fact that $h_\ep$ satisfies
\beq\label{eqhep}
\frac{1}{\he} \( - \div \(\frac{\nab h_\ep}{a_\ep}\) +h_\ep \) = \mu_\ep +\psi_\ep\end{equation}
 where  $\psi_\ep$ tends to 0  and $\mu_\ep$ defined in (\ref{23}) tends to some $\mu_0$, both convergences being strong  in $W^{-1,r}$ for $r<2$. The notion of $H$-convergence and a priori estimates allow us to pass to the limit in (\ref{eqhep}) in order to  get that the weak  $H^1$  limit of $h_\ep /\he$, that we call $h_0$, solves 
\begin{equation}
\label{eqh0}
-\div (\A\nab h_0 ) + h_0 = \mu_0.\end{equation}
Combining the lower bounds of the energy inside and outside the vortex balls (\ref{24})-(\ref{25}), we find
$$\liminf_{\ep\to 0}\frac{1}{\he^2}J_\ep(u_\ep,A_\ep)\ge E(h_0) \ge E(h_*).$$
The last inequality is true because (\ref{eqh0}) implies that $h_0$ is in $V$.

Section III is devoted to the proof of Proposition \ref{propupperbd}. The proof holds for any positive Radon measure $\mu$. We apply it to $\mu_*$ to get that
$$\limsup_{\ep\to 0} \frac{1}{\he^2} J_\ep(u_\ep,A_\ep)\leq E(h_*),$$
which will imply the desired results of convergence.

The upper bound of  Proposition \ref{propupperbd} is obtained by constructing test configurations as follows. First, given a positive Radon measure $\mu$, we construct approximate measures $\mue$ which  converge weakly to $\mu$:
$$\mue = \frac{1}{\he} \sum_{i=1}^{n_\ep}\mie,$$
where $\mie$ is the line element on the circle $\p B(\pie,\ep)$ normalized so that $\mie(\p B(\pie,\ep)) = 2\pi$. The measure $\mue$ describes the vortices of our test-configuration.
The difficulty is to choose the points $\pie$ satisfying a number of properties. We tile $\Omega$ with squares $K$ of size $\delta(\ep)$. In each square, there is at least a point $p_K$ where $\beta_\ep = 0$. We choose $n_\K$ points $\pie$
 regularly scattered around $p_K$ in a ball of radius $1/\he$. The number $n_K$ is chosen depending on $\mu(K)$ so that $\mue$ converge  to $\mu$.
  Once  the vortices are constructed, the rest follows easily: the magnetic field $h_\ep$ is defined to be the solution of 
\beq\label{eqhepupper}
\frac{1}{\he} \( - \div \(\frac{\nab h_\ep}{a_\ep}\) +h_\ep \) = \mu_\ep .\end{equation}
Then, we are the able to construct a configuration $(u_\ep,A_\ep)$ such that   $\curl A_\ep=h_\ep$ and $u_\ep$ has vortices at the points $\pie$. Moreover, we obtain
$$ J_\ep(u_\ep,A_\ep) \approx \frac{1}{2} \io \frac{1}{a_\ep} |\nabla h_\ep |^2+|h_\ep -1|^2.$$
Finally we are able to show that
$$ \limsup_{\ep\to 0}\frac{1}{2\he^2} \io \frac{1}{a_\ep} |\nabla h_\ep |^2+|h_\ep -1|^2 \le  \frac{\La}{2}\io b \ d\mu +\hal\io \nabla h \cdot \A \nabla h +|h-1|^2,$$
where $h$ solves $ - \div (\A\nab h) +h = \mu$ and $h=1$ on $\bo$.


\section{Lower bound}

In the following, we will denote $\nabla_A u=\nabla u -iAu$. We will often drop the subscripts $\ep$. We consider $(u_\ep, A_\ep)$ a family of minimizers of $J_\ep$, thus a family of solutions of (G.L.). We can state a few a priori bounds. Firstly, by the maximum principle, $|u_\ep|\le \max a_\ep \le 1$. Secondly, by minimality, comparing with $(a_\ep, 0)$, we get
$$J_\ep(u_\ep,A_\ep) \le J_\ep(a_\ep, 0).
$$
But, by hypothesis (H2) on $a_\ep$, $$J_\ep(a_\ep, 0) = \hal \io|\nab a_\ep|^2 +O(\he^2) \le \frac{C}{\eta^2}+O(\he^2) \le C\he^2 .$$ 
Hence, we have the a-priori estimate 
\begin{equation} 
\label{I1}
J_\ep(u_\ep,A_\ep) \le C\he^2.
\end{equation}
In addition, by applying a gauge-transformation to $(u_\ep, A_\ep)$, we can choose the Coulomb-gauge $\div A_\ep =0$ in $\om$, with $A_\ep . n=0$ on $\bo$. With this choice of gauge, we are easily lead (see \cite{s1,ss1}) to the a priori bounds 
\begin{eqnarray}
\label{apb1}
\|A_\ep\|_{L^\infty(\om)} &\le C \he\\
\label{apb2}
\|\nab u_\ep\|_{L^2(\om)} & \le C\he.\end{eqnarray}
We begin with the proof of Proposition \ref{propballs}.
\subsection{Proof of Proposition \ref{propballs}}
{\it - Step 1 :}
Let $(u,A)$ be an energy-minimizer. Denoting  $|u|$ by $\ro$, since $\io |\nab u|^2 \ge \io|\nab \ro|^2$, we deduce from (\ref{I1}) :
\begin{equation}
\label{I2}
\io|\nab \ro|^2 +\frac{1}{2\ep^2}(\ro^2-a_\ep)^2 \le C\he^2. \end{equation}
But, 
\begin{eqnarray*}
\io|\nab \ro|^2 & = & \io|\nab (\ro -\sqrt{a_\ep})|^2 +|\nab \sqrt{a_\ep}|^2 - 2 \nab (\ro-\sqrt{a_\ep})\cdot\nab \sqrt{a_\ep}\\
& \ge & \io |\nab (\ro -\sqrt{a_\ep})|^2- 2 |\nab (\ro-\sqrt{a_\ep})| |\nab \sqrt{a_\ep}|.\end{eqnarray*}
Hence, in view of (\ref{I2}),
\begin{eqnarray*}
\io|\nab (\ro-\sqrt{a_\ep})|^2 & \le  & C\he^2 + \|\nab (\ro-\sqrt{a_\ep}) \|_{L^2} \|\nab \sqrt{a_\ep}\|_{L^2}\\
&\le&C\he^2+\frac{C}{\eta(\ep)}\|\nab(\ro-\sqrt{a_\ep})\|_{L^2}
, \end{eqnarray*}
and, since $\frac{1}{\eta(\ep)}\ll \he$, 
$$\io|\nab (\ro -\sqrt{a_\ep})|^2 \le \max(C\he^2, \frac{C}{\eta^2} ) \le C\he^2.$$ 
In view of (\ref{I2}), we thus have
\begin{equation}
\label{I3}
\hal\io|\nab (\ro - \sqrt{a_\ep} )|^2 +\frac{1}{2\ep^2} (a_\ep-\ro^2)^2 \le C \he^2 \le C\lep^2.\end{equation}
\noi
{\it - Step 2 :} For any $t\in \mr$, let $\om_t= \{x\in \om/ |\ro- \sqrt{a_\ep}|(x) >t\} $ and $\gamma_t = \p \om_t.$ 
Applying the coarea formula and arguing as in Lemma IV.2 of \cite{ss2},
\begin{eqnarray*}
C\lep^2 \ge \io|\nab (\ro - \sqrt{a_\ep})|^2 +\frac{1}{2\ep^2}(a_\ep- \ro^2)^2 &  \ge &  \frac{C}{\ep}\io|\nab (\ro - \sqrt{a_\ep})||a_\ep-\ro^2|\\
& \ge & \frac{C}{\ep} \int_0^{+\infty} r(\gamma_t)t dt.
\end{eqnarray*}
Here, as in \cite{ss2}, $r(\gamma_t)$ is defined as the infimum over all finite coverings of $\gamma_t$ by balls $B_1, \cdots, B_k$ of the sum $r_1 + \cdots+ r_k$ where $r_i$ is the radius of $B_i$.
Combining the previous inequality with  the mean-value theorem, we find that there exists a $t\in \left[0, \frac{1}{\lep}\right]$ such that $r(\gamma_t) <C\ep \lep^3.$

\hfill

\noi
{\it - Step 3 : }The next step is to construct the vortex-balls : starting from the chosen $\gamma_t$, covered by balls $B_1, \cdots, B_k$ (whose sum of the radii is controlled by $C\ep \lep^3$), we use the method of growing and merging of balls used in \cite{sa, ss2} : one needs to grow these balls $B_i$, keeping a suitable lower bound on the energy they contain, until the desired size is reached, with the desired lower bound.  
When some balls happen to intersect during the growth process, they are merged into a larger one. We refer the reader to \cite{ss2}, and here we only need to apply the result of Proposition IV.1 of \cite{ss2} to $A_\ep$ and
$v = \frac{u}{|u|}= e^{i \varphi}$ in $ \om \backslash \om_t$, $\sigma = e^{-\sqrt{\lep}}$. We then obtain the existence of balls $B_i=B(p_i, r_i)$ such that (\ref{usmall}) and (\ref{sumri}) hold, and 
\begin{equation}
\label{I4}
\hal \int_{B_i \backslash \om_t} |\nab \varphi-A|^2 +\hal \int_{B_i} |h-\he|^2 \ge \pi |d_i|\lep(1-o(1)), 
\end{equation}
with $d_i = \text{deg}(u, \p B_i)$ if $\overline{B_i} \subset \om$, and $0$ otherwise. But we also have, from the Ginzburg-Landau equation $-\np h = \ro^2(\nab \varphi-A)$, and from $\ro  \le 1$,
$$\io|\nab h|^2 = \io\ro^4|\nab \varphi-A|^2 \le \io|\nab_A u|^2 \le C\he^2, 
$$hence 
\begin{eqnarray*}
\int_{B_i} |h-\he|^2&\le& C r_i \|h-\he\|_{L^4(\om)}^2  \le C r_i \|h-\he\|_{H^1(\om)}^2 \\
& \le &  C\he^2 e^{-\sqrt{\lep}}  = o(1). \end{eqnarray*}
Thus, (\ref{I4}) becomes
\begin{equation}
\label{I5}
\hal \int_{B_i \backslash \om_t}|\nab \varphi-A|^2  \ge \pi |d_i| \lep(1-o(1)). \end{equation}
Now,
\begin{eqnarray*}
\hal \int_{B_i \backslash \om_t} |\nab_A u|^2 & \ge & \hal \int_{B_i \backslash \om_t}\ro^2
|\nab \varphi-A|^2
\\
& \ge & \hal  \int_{B_i \backslash \om_t}a_\ep
|\nab \varphi-A|^2+ \hal \int_{B_i \backslash \om_t}(\ro^2 - a_\ep)|\nab \varphi-A|^2
\\& \ge & 
\hal \( \min_{B_i} a_\ep\) \int_{B_i \backslash \om_t}|\nab \varphi-A|^2 -\frac{C}{\lep} \int_{B_i \backslash \om_t}|\nab \varphi-A|^2, \end{eqnarray*}
where we have used (\ref{usmall}). 
In view of (\ref{I5}),
$$\hal \int_{B_i \backslash \om_t} |\nab_A u|^2  \ge   \pi\( \min_{B_i} a_\ep\) |d_i|\lep (1-o(1)). $$
So, using the hypotheses (H2) and (H3) on $a_\ep$, we are led to the two following lower bounds
\begin{eqnarray}
\label{lb1}
\hal \int_{B_i \backslash \om_t} |\nab_A u|^2 &\ge  &   \pi a_\ep(p_i) |d_i|\lep(1-o(1))\\
\label{lb2}
\hal \int_{B_i \backslash \om_t} |\nab_A u|^2 & \ge &  \pi b(p_i) |d_i| \lep (1-o(1)). \end{eqnarray}
This proves (\ref{intbi}).
\hfill $\Box$

\subsection{Deriving the limiting equation}
For any $(p_i, d_i)$ satisfying (\ref{usmall})---(\ref{intbi}), we can define
\begin{equation}
\label{I5b}  
\mu_\ep = \frac{2\pi}{\he} \sum_{i\in I_\ep} d_i \delta_{p_i},\end{equation}
a measure of vorticity per unit of applied field. We will see that it remains a  bounded family of measures.
\begin{lem}
If $\Lambda>0$, and $(u_\ep,A_\ep)$ is a family of minimizers of $J_\ep$ with $h_\ep = \curl A_\ep$, we can extract a sequence $\ep_n \to 0$ such that there exists $h_0-1  \in H^1_0(\om)$, and $\mu_0 \in \mathcal{M} $ with 
$$\frac{h_{\ep_n} }{\he} -1 \rightharpoonup h_0-1  \quad \text{ in } H^1_0(\om), $$
$$ \mu_{\ep_n} \to \mu_0 \quad\text{in the sense of measures}. $$
\end{lem}
{\it Proof :} As seen in the previous proof, since $(u_\ep,A_\ep)$ is a solution of the second Ginzburg-Landau equation
$$ \io |\nab h_\ep |^2 \le \io|\nab_{A_\ep} u_\ep |^2 \le C \he^2 $$ and 
$$\io |h_\ep - \he |^2 \le C \he^2 .$$
Hence, $\frac{h_\ep}{\he} - 1$  is bounded in $H^1_0(\om)$, and we can find  a sequence $\ep_n \to 0$ such that $\frac{h_{\ep_n}}{\he}$ converges weakly in $H^1_0$ to some $h_0-1$. On the other hand, from Proposition \ref{propballs}, 
\begin{eqnarray*}
C \he\frac{\lep}{\Lambda}   \ge  J_\ep(u_\ep,A_\ep) & \ge & \sum_{i \in I_\ep} \pi |d_i|b(p_i) \lep (1-o(1)) \\
& \ge & b_0 \sum_i \pi |d_i| \lep(1-o(1)), \end{eqnarray*}
where $b_0$ is given by hypothesis (H1) on $a_\ep$. 
Hence, $$\hal\io |\mu_{\ep_n} | = \frac{\pi \sum_i |d_i|}{\he} \le C, $$ thus 
$(\mu_{\ep_n} )$  is a bounded sequence of measures, and extracting again if necessary, we can assume that $\mu_{\ep_n} $ converges to some $\mu_0$ in the sense of measures. 
\hfill $ \Box$

\hfill

\begin{pro}
Let $\mu_0$ and $h_0$ be the measures and fields defined in Lemma II.1.  Then there exists $r_0<2$ such that $\mu_0\in W^{-1,r}(\om)$ $\forall r\in (r_0,2)$, and  $h_0$ is  the unique solution in $W^{1,r}$ of 
\begin{equation}
\label{I6}
\left\{\begin{array}{ll}
-\div \(\A\nab h_0\) +h_0 = \mu_0 & \text{ in }  \om
\\
h_0 = 1  & \text{ on }  \bo.\end{array}
\right.
\end{equation}
\end{pro}
The proof of this proposition requires the following lemma, a slight refinement of the result stated in \cite{ss1}, Lemma II.3.
\begin{lem} 
Under the  hypotheses of Lemma II.1, for any $q>2$,
$$\frac{1}{\he} \curl \frac{(iu_\ep, \nab u_\ep)}{a_\ep} - \mu_{\ep} \mathop{\longrightarrow}_{\ep \to 0} 0 \quad \text{ strongly in } (W_0^{1,q}(\om) )'. $$ 
\end{lem}
{\it Proof : } Denote $\tilde{\om} = \om \backslash \cup_i B_i $. On $\tilde{\om}$, $ |u_\ep|\ge b_0>0$ and $v_\ep= \frac{u_\ep}{|u_\ep|}$ is well-defined. Let $q>2$, and  $\xi \in W_0^{1,q}$. 
We need to show that 
$$\left|\frac{1}{\he} \io \xi\curl\frac{(iu_\ep, \nab u_\ep)}{a_\ep} -\frac{2\pi}{\he} \sum_i d_i \xi(p_i)\right|\le o(1) \|\xi\|_{W^{1,q}_0(\om)}
.$$
Dropping again some of the subscripts, we have
\begin{equation}
\label{I7}
\frac{1}{\he} \io \xi \curl\frac{(iu, \nab u)}{a_\ep} = - \frac{1}{\he} \io \np \xi \cdot\frac{(iu, \nab u)}{a_\ep}. 
\end{equation}
Then, the method consists in splitting this integral into the integral over the vortex-balls (which is going to be negligible because the balls are small enough) and the integral over $ \tilde{\om}$, the complement of the balls.

\hfill

\noi
{\it - Step 1 :}
We prove that 
\begin{equation}
\label{I8}
\left|\ \int_{\cup_i B_i} \frac{1}{\he} \np \xi \cdot\frac{(iu, \nab u)}{a_\ep}\right| = o(1) \|\nab \xi\|_{L^q(\om)}. \end{equation}
Indeed, since $a_\ep \ge b_0>0$, 
$$ \left|\ \int_{\cup_i B_i} \frac{1}{\he} \np \xi \cdot\frac{(iu, \nab u)}{a_\ep}
\right|  \le \frac{1}{b_0} \frac{\|\nab u \|_{L^2(\om)} } {\he} \|\nab \xi\|_{L^q} (\text{vol}(\cup_i B_i ))^{\frac{1}{p}}, $$
where $\frac{1}{p}+ \frac{1}{q} = \hal $ and we have used H\"older's inequality twice. Using (\ref{apb2}),
$$\left| \ \int_{\cup_i B_i} \frac{1}{\he} \np \xi \cdot\frac{(iu, \nab u)}{a_\ep}\right| \le C (\sum_i r_i ^2)^{\frac{1}{p}} \|\nab \xi\|_{L^q(\om) }.$$
In addition, $(\sum_i r_i ^2)^{\frac{1}{p}} \le (\sum_i r_i)^{\frac{2}{p}} = o(1) $ since we know that $\sum_i r_i  \to 0 $.
Therefore, (\ref{I8}) is proved.

\hfill

\noi
{\it - Step 2 : } We observe that \begin{eqnarray}
\nonumber
\frac{1}{\he} \int_{\tilde{\om}}\np \xi \cdot\frac{(iu, \nab u)}{a_\ep}
& = & \frac{1}{\he} \int_{\tilde{\om}}\frac{|u|^2}{a_\ep} (iv, \nab v) \cdot \np \xi
\\
& = & \frac{1}{\he}  \int_{\tilde{\om}}(iv, \nab v)\cdot \np\xi +\frac{1}{\he} 
\int_{\tilde{\om}} \left(\frac{|u|^2}{a_\ep} -1\right) (iv, \nab v)\cdot\np \xi. \end{eqnarray}We claim that
\begin{equation}
\label{I10}
\frac{1}{\he}\left|\int_{\tilde{\om}} \left(\frac{|u|^2}{a_\ep} -1\right) (iv, \nab v)\cdot\np \xi\right|\le o(1) \|\nab \xi\|_{L^q}
.\end{equation}
Indeed, 
\begin{eqnarray*}
\frac{1}{\he}\left|\int_{\tilde{\om}} \left(\frac{|u|^2}{a_\ep} -1\right) (iv, \nab v)\cdot\np \xi\right| & \le & \frac{1}{b_0 \he} \left|\int_{\tilde{\om}}(|u|^2 - a_\ep ) |\nab v||\nab \xi|\right|\\
& \le & C\frac{\|\nab v\|_{L^2(\tilde{\om})}}{\he} \|\nab \xi\|_{L^q(\om)} \| |u|^2 - a_\ep \|_{L^p(\om)}, \end{eqnarray*}
with $\frac{1}{p}+\frac{1}{q}= \hal$. From the a priori estimate (II.1), 
$$ \io (|u|^2 - a_\ep)^p \le C\io (|u|^2-a_\ep)^2 \le C \ep^2 \he^2 = o(1),$$
hence, using $\|\nab v\|_{L^2(\tilde{\om})} \le C \|\nab u\|_{L^2(\om) } \le C \he$,  we obtain (\ref{I10}).
Combining (\ref{I7})---(\ref{I10}), we have 
\begin{equation}
\label{I11}
\frac{1}{\he} \io \curl\frac{(iu, \nab u)}{a_\ep} \xi = \frac{1}{\he}\int_{\tilde{\om}} (iv, \nab v) \cdot \np \xi + o(1) \|\xi\|_{W_0^{1,q}}. \end{equation}

\hfill

\noi
{\it - Step 3 :}
We evaluate $\D \int_{\tilde{\om}} (iv, \nab v)\cdot \np\xi.$
Noticing that $\curl (iv, \nab v)\equiv 0 $ on $\tilde{\om}$, we have
$$\int_{\tilde{\om}} (iv, \nab v)\cdot \np\xi = \int_{\p \tilde{\om}} \xi \(iv,\frac{\p v}{\p \tau}\) = \sum_i \int_{\p B_i\cap \om} \xi  \(iv,\frac{\p v}{\p \tau}\).$$
There remains to prove that 
\begin{equation}
\label{sumxi}
\sum_i \int_{\p B_i\cap \om} \xi  \(iv,\frac{\p v}{\p \tau}\)=
2\pi \sum_i d_i \xi(a_i) +o(\he) \|\xi\|_{W^{1,q}_0(\om)}. \end{equation}
Let $f$ be a $C^1$ function defined on $\mr_+$ such that
\begin{equation}
\label{deff}
\left\{ \begin{array}{ll}
f(x)=x & \text{ for } x\le \frac{b_0}{2}\\
f(x)=1 & \text{ for } x \ge b_0\\
|f'(x)|\le C & \text{ for any } x \ge 0. 
\end{array}
\right.
\end{equation}
We can define the complex-valued function
\begin{equation}
\label{defw}
w = f(|u|) v. \end{equation}
It has a meaning everywhere by setting $w=u$ where $|u|\le \frac{b_0}{2}$.
Then, it is easy to check that 
\begin{equation}
\label{gradw}
|\nab w|\le C |\nab u| \quad \text{ in } \om, 
\end{equation}
and 
\begin{equation}
\label{devaw}
\sum_i \int_{\p B_i\cap \om} \xi  \(iv,\frac{\p v}{\p \tau}\)=
\sum_i \int_{\p B_i\cap \om} \xi  \(iw,\frac{\p w}{\p \tau}\). \end{equation}
Using Stokes theorem, we have
\begin{equation}
\label{cinq}
\left|\sum_i \int_{\p B_i}(\xi - \xi(p_i)) \( iw, \frac{\p w}{\p \tau}\)\right|  = \left|\sum_i \int_{B_i} \np \xi\cdot (iw, \nab w) + (\xi-\xi(p_i)) \curl (iw, \nab w)\right|.
\end{equation}
But, on the one hand,
\begin{eqnarray}
\label{I15}
\nonumber 
 \frac{1}{\he}\left|\sum_i \int_{b_i} \np \xi\cdot (iw, \nab w)\right| & \le & C\frac{\|\nab w\|_{L^2} }{\he} \|\nab \xi\|_{L^q}  \(\sum_i \text{vol} (B_i)\) ^{\frac{1}{p}}\\
\nonumber & \le & C\frac{\|\nab u\|_{L^2} }{\he} \|\nab \xi\|_{L^q}  \(\sum_i r_i^2\) ^{\frac{1}{p}}\\
& \le & o(1) \|\nab \xi\|_{L^q} 
\end{eqnarray}
as in the proof of (\ref{I8}). On the other hand, using the fact that, since $q>2$, $W_0^{1,q}$ embeds in  $C^{0, \beta}$ for some $\beta <1$, and $|\curl (iw, \nab w)| \le C |\nab w|^2\le C |\nab u|^2,$ we have
\begin{eqnarray}
\label{I16}
\nonumber
\left|\sum_i \frac{1}{\he}\int_{\p B_i}(\xi - \xi(p_i)) \curl (iw, \nab w) \right| & \le & (\max_i  r_i)^{\beta} \|\xi\|_{C^{0, \beta}(\om)}  \sum_i \int_{U_i}  \frac{|\nab u|^2}{\he} \\
\nonumber
& \le & e^{-\beta \sqrt{\lep}} \frac{\|\nab u\|_{L^2(\om)}^2 }{\he} \|\xi\|_{W_0^{1,q}}\\
& \le & \he   e^{-\beta \sqrt{\lep}}\|\xi\|_{W_0^{1,q}} = o(1) \|\xi\|_{W_0^{1,q}}, \end{eqnarray}
since $\he \le C \lep$.
As in \cite{ss1}, the proof remains valid even if $B_i$ intersects $\bo$. Combining (\ref{I15}),  (\ref{I16}), (\ref{devaw}),  and (\ref{cinq}), (\ref{sumxi}) is proved. Consequently, in view of (\ref{I11}), we can conclude that 
$$\left|
\frac{1}{\he} \io\xi \curl \frac{(iu, \nab u)}{a_\ep}  - \frac{2\pi}{\he} \sum_i d_i \xi(p_i) \right|\le o(1) \|\xi\|_{W_0^{1,q}}, $$
hence that $\frac{1}{\he}\curl \frac{(iu, \nab u)}{a_\ep} - \mu_\ep\to 0 $ strongly in $(W_0^{1,q})'$ as stated.
\hfill $\Box$

\hfill

\noi
{\it Proof of Proposition II.1 :} For the sake of simplicity, we write $\ep$ instead of $\ep_n$. 

\hfill

\noi
{\it - Step 1 :} We prove that $h_\ep$ satisfies
\begin{equation}
\label{I17}
\frac{1}{\he} \( - \div \(\frac{\nab h_\ep}{a_\ep}\) +h_\ep \) = f_\ep
,\end{equation}
with $f_\ep = \mu_\ep + \psi_\ep$, where $\psi_\ep \to 0 $ strongly in $(W_0^{1,q})'$  for $q>2$. Indeed, we start from the second Ginzburg-Landau equation :
$$ - \np h_\ep = (iu_\ep, \nab_{A_\ep} u_\ep),$$
divide it by $a_\ep$ and take the curl : 
$$ 
- \div \(\D\frac{\nab h_\ep}{a_\ep}\) = \curl\( \D\frac{(iu_\ep, \nab u_\ep )}{a_\ep} - A_\ep\frac{|u_\ep|^2}{a_\ep} \),$$
hence 
\begin{equation}
\label{I18}
- \div \(\D\frac{\nab h_\ep}{a_\ep}\) +h_\ep =\curl \D\frac{(iu_\ep, \nab u_\ep )}{a_\ep}+ \curl \(A_\ep \(1-\D\frac{|u_\ep|^2}{a_\ep} \)\).
\end{equation}
Now consider a test-function $\xi\in W_0^{1,q}(\om)$, $q>2$,
\begin{eqnarray*}
\left|\io \xi \curl \(A_\ep \(1-\frac{|u|^2}{a_\ep} \)\)\right|  & = & \left|\io \np \xi\cdot A_\ep\( 1- \frac{|u|^2}{a_\ep} \)\right|\\
& \le & C \|A_\ep\|_{L^{\infty}(\om)} \|\nab \xi\|_{L^2(\om)}\|a_\ep -|u|^2 \|_{L^2(\om)}.\end{eqnarray*}
The a-priori bound (\ref{apb1}),  $\|A_\ep\|_{L^\infty(\om)}\le O(\he)$ and the energy bound, $\|a_\ep -|u|^2\|_{L^2 } \le C \ep\he,$
 yield
$$\left|\io \xi \curl \(A_\ep \(1-\frac{|u|^2}{a_\ep} \)\)\right| \le o(1)\|\nab \xi\|_{L^2}
.$$
Consequently, $\curl \(A_\ep \(1-\frac{|u|^2}{a_\ep} \)\) \to 0$ strongly in $(W_0^{1,q})'$  for $q>2$. Combining this with (\ref{I18}) and Lemma II.2, we get the desired result.

\hfill

\noi
{\it - Step 2 :} We prove that $f_\ep$ converges to $\mu_0$, the weak limit of $\mu_\ep$, in $W^{-1, r}(\om)$ for any $r<2$. Indeed, from the upper bound on the energy, we know that $\frac{1}{a_\ep \he} \nab h_\ep$  is bounded  in $L^2(\om)$, hence, in view of (\ref{I17}),  $f_\ep$ is bounded in $H^{-1}$, hence in $W^{-1,p}$ for $p<2$. But, on the other hand, $f_\ep = \mu_\ep +\psi_\ep$, with $\psi_\ep$ bounded in $W^{-1, p}$ for $p<2$, hence $\mu_\ep$ remains bounded in $W^{-1,p}$ for $p<2$. 
Furthermore, $\mu_\ep$ is also bounded in the sense of measures, therefore we can apply a theorem of Murat (see \cite{mu1}) which asserts that such a $\mu_\ep$, bounded in the sense of measures and in $W^{-1,p}$ for $p<2$, is necessarily compact in $W^{-1,r}$ for $r<p$. Since this is also the case for $\psi_\ep$, which converges to zero, this implies that $f_\ep$ is compact in $W^{-1,r}$ for $r<2$. In addition, its limit in the sense of distributions is $\mu_0$, hence it must converge to $\mu_0$ in $W^{-1,r}$.

\hfill

\noi
{\it - Step 3 :} We wish to pass to the limit in (\ref{I17}), but it is not possible directly because the $H$-convergence requires a right-hand side in $H^{-1}$. So we are going to pass to the limit in the duality sense for a fixed right-hand side.
 Let $g\in W^{-1,q}$ for $q>2$. 
 Using the hypothesis (H1) on $a_\ep$, (which implies in particular the uniform ellipticity of $\frac{1}{a_\ep  }\Id$), we can apply a  theorem of Meyers \cite{me} : there exists a $q_0>2$, such that  if $g$ is in $W^{-1,q}$ with $2<q\le q_0$, then equation 
\begin{equation}
\label{I19}
\left\{\begin{array}{ll}
-\div \(\D\frac{\nab v_\ep}{a_\ep}\) +v_\ep = g & \text{ in }  \om
\\
v_\ep = 0  & \text{ on }  \bo,\end{array}
\right.
\end{equation}
 has a unique solution $v_\ep$ in $W^{1,q}_0$. Thus, we have 
\begin{equation}
\label{I21}
_{W^{1,q'}_0} <\frac{h_\ep}{\he}-1 ,g>_{W^{-1,q}}= _{W^{-1,q'}}<f_\ep-1 , v_\ep>_{W^{1,q}_0},
\end{equation}
where $\frac{1}{q'}+\frac{1}{q}=1$, and we want to pass to the limit.

More precisely, Meyers' theorem yields that the operator $R_\ep$ which maps $g$ to $v_\ep$, is  a bounded linear operator from $W^{-1,q}$ to $W^{1,q}_0$ (for $2<q\le q_0$), hence up to extraction of a subsequence, $v_\ep$  has a weak limit $v_0$ in $W^{1,q}_0$.
 We assumed in hypothesis (H4) that $\frac{1}{a_\ep} \Id $ $H$-converges to $\A$. By  the definition of $H$-convergence
(see \cite{mut}), and since $W^{1,q}_0\subset H^1_0$, this implies that $v_0$ is the  solution of 
\begin{equation}
\label{I22}
\left\{\begin{array}{ll}
-\div \(\A\nab v_0\) +v_0 = g & \text{ in }  \om
\\
v_0 = 0  & \text{ on }  \bo,\end{array}
\right.
\end{equation}
Since this possible weak limit $v_0$ is unique, the whole sequence $v_\ep$ converges to $v_0$ weakly  in $W^{1,q}_0$.
In addition, $f_\ep$ converges strongly to $\mu_0$ in $W^{-1, q'}$, thus we have
$$ _{W^{-1,q'}}<f_\ep-1 , v_\ep>_{W^{1,q}_0} \to <\mu_0-1 , v_0>.$$
On the other hand, $\frac{h_\ep}{\he}-1$ converges weakly to $h_0-1$ in  $H^1_0$. Thus, 
$$
_{W^{1,q'}_0} <\frac{h_\ep}{\he}-1, g>_{W^{-1,q}} \to <h_0-1, g>.$$
Therefore, we can pass to the limit in (\ref{I21}), and we are led to
\begin{equation}
\label{I23}
_{W^{1,q'}_0}<h_0-1 , g>_{W^{-1,q}}= _{W^{-1,q'}}<\mu_0-1 , v_0>_{W^{1,q}_0}.\end{equation}
Meyers' aforementioned theorem, also yields that for $q_0'\leq q' <2$, (\ref{I6}) has a unique solution in $W^{1,q'}$. Since (\ref{I23}) holds for any $g$ in $W^{-1,q}$, it implies that
 $h_0$ is this solution.
\hfill $ \Box$

\subsection{Deriving a lower bound outside the vortex balls}
Next, we would like to deduce from (\ref{I6}) a lower bound like
$$\liminf_{\ep\to 0} \int_{\om \backslash \cup_i B_i} \frac{|\nab h|^2}{a_\ep\he^2} \ge \io \nab h_0 \cdot \A \nab h_0.$$
But this is impossible to derive straightforwardly because the domain of integration in the left-hand side integral is not $\om$. To remedy this, we replace $h_\ep$ by an auxiliary field $\hb$, a sort of truncated of $h_\ep$ in the balls. This is  a trick that was already used in \cite{ss2} Proposition IV.1, Step 1.
\begin{lem}
There exists $\hb$ such that $\hb-1 \in H^1_0(\om)$ and   

\noi
1) $\frac{\hb}{\he}-1  \rightharpoonup h_0-1 $ in $H^1_0(\om)$,\\
2) $$\int_{\om\backslash \cup_i B_i}
\frac{|\nab h|^2}{a_\ep} +\io |h_\ep - \he|^2 \ge \io \frac{|\nab \hb|^2}{a_\ep}+|\hb -\he|^2 -o(1),$$
3) $$\liminf_{\ep \to 0 } \io \frac{| \nab \hb|^2}{a_\ep} \ge \io \nab h_0 \cdot \A \nab h_0.$$

\end{lem}
{\it Proof :} We consider $\overline{A_\ep}$  a solution of the following minimization problem :
\begin{equation}
\label{I25}
\mathop{\min}_{A\in H^1(\om, \mr^2), \text{div }  A =0}  \int_{\Omega\backslash \cup_i B_i} a_\ep |\nab \varphi - A|^2 +\io |\curl A -\he|^2 , \end{equation}
where $\nab \varphi$ denotes the gradient of the phase of $u_\ep$ which is well-defined in $\om\backslash \cup_i B_i$. If we write $\hb= \curl \overline{A_\ep}$, and we test  (\ref{I25}) with $h_\ep$, we have
\begin{equation}
\label{I26}
\int_{\om\backslash \cup_i B_i } a_\ep |\nab \varphi- \overline{A_\ep} |^2 +\io|\hb - \he|^2 \le \int_{\om\backslash \cup_i B_i } a_\ep |\nab \varphi- A_\ep |^2+\io |h_\ep - \he|^2 \le C \he^2. \end{equation}
In addition,  $\hb $ and $\overline{A_\ep}$ satisfy the following equations :
\begin{equation} 
\label{I27}
\left\{\begin{array}{ll}
-\np \hb = a_\ep (\nab \varphi- \overline{A_\ep}) & \text{ in }\om\backslash \cup_i B_i\\
\hb= cst= c_i & \text{ on } B_i, \forall i\\
 \hb = \he & \text{ on }\bo.\end{array} 
\right. \end{equation}
Thus, it satisfies 
\begin{equation}
\label{I28}
- \div \( \frac{\nab \hb}{a_\ep \he}\) +\frac{\hb}{\he} = \nu_\ep, \end{equation}
where $\nu_\ep$ is the measure defined by 
\begin{equation}
\label{I29}
\forall \xi\in W^{1,q}_0(\om), (q>2),\quad 
\io \nu_\ep \xi = \sum_i \frac{1}{\he} \int_{\p B_i} \xi\frac{\p \varphi}{\p \tau} 
+ \sum_i \frac{1}{\he} \int_{B_i} c_i \xi.\end{equation}
On the other hand, using Cauchy-Schwartz inequality,
$$\left|\frac{1}{\he} \sum_i \int_{B_i} c_i \xi \right|= \left| \frac{1}{\he} \int_{\cup_i B_i} \hb \xi\right|\le \|\xi\|_{L^\infty}\left\|\frac{\hb}{\he}\right\|_{L^2} \(\sum_i r_i\)^{\hal}.$$
In view of (\ref{I26}), $\left\|\frac{\hb}{\he}\right\|_{L^2}$ is bounded, and $(\sum_i r_i)^\hal\le \sum_i r_i \to 0 $ from Proposition \ref{propballs}.
Hence, 
$$\left|\frac{1}{\he} \sum_i \int_{B_i} c_i \xi \right|=o(1)\|\xi\|_{L^\infty}
.$$
On the other hand, the same proof as for Lemma II.2 shows that
$$\left|\sum_i \frac{1}{\he} \int_{\p B_i} \frac{\p \varphi}{\p \tau} \xi - \io \xi d\mu_\ep\right|= o(1) \|\xi\|_{W^{1,q}_0}.$$
Hence, in view of (\ref{I29}), $\nu_\ep - \mu_\ep $ converges strongly to $0$ in $(W^{1,q}_0)'$. The same argument as in Proposition II.1 allows to conclude from (\ref{I28}) that $$\frac{\hb}{\he}-1 \rightharpoonup h_0-1 \quad \text{ in } H^1_0(\om),$$
using the uniqueness of the solution of (\ref{I6}).

Using (\ref{I26}) and (\ref{I27}), we get
\begin{eqnarray*}
\io \frac{|\nab \hb|^2}{a_\ep} + |\hb -\he|^2  & = & \int_{\om\backslash\cup_i B_i} a_\ep |\nab \varphi- \overline{A_\ep}|^2 +\io |\hb - \he|^2 \\
& \le & \int_{\om\backslash\cup_i B_i} a_\ep |\nab \varphi-A_\ep|^2 +\io |h_\ep - \he |^2 
.\end{eqnarray*}
As in the proof of Proposition \ref{propballs}, we have
$$\int_{\om\backslash\cup_i B_i} a_\ep |\nab \varphi- A_\ep|^2 \le 
\int_{\om\backslash\cup_i B_i}\frac{|\nab h_\ep|^2}{a_\ep} +o(1).$$
Thus, assertion 2) is proved. In addition, $\frac{\hb}{\he}-1$ is bounded in $H^1_0(\om)$ and the convergence to $h_0-1$ is weak in $H^1_0$. There remains to prove the third assertion. But it is a classical result in homogenization theory (see \cite{jo}) that, since $\frac{\hb}{\he}-1 \rightharpoonup h_0-1$ in $H^1_0(\om)$ and $\frac{1}{a_\ep}\Id $ $H$-converges to $\A$, 
$$\liminf_{\ep \to 0} \io\frac{1}{a_\ep} \left|\nab \(\frac{\hb}{\he}\)\right|^2 \ge \io \nab h_0 \cdot \A \nab h_0. $$
This completes the proof of the lemma.
\hfill $\Box$

\hfill

\noi
We recall that we defined $E$ in (\ref{E(f)}). 
\begin{lem}
With the same notations, 
$$\liminf_{\ep \to 0} \frac{J_\ep(u_\ep,A_\ep)}{\he^2} \ge \frac{\Lambda}{2}\io b |\mu_0|+\hal\io \nab h_0\cdot  \A \nab h_0+|h_0-1|^2 =E(h_0).$$
\end{lem}
{\it Proof :} The energy can  easily be bounded from below as follows, splitting between the contribution inside the vortex-balls and the contribution outside :
\begin{eqnarray*}
J_\ep(u_\ep,A_\ep) & \ge & \hal \io |\nab_A u|^2 +|h-\he|^2\\
& \ge & \hal \int_{\cup_{i \in I} B_i} |\nab_A u|^2 +\hal \int_{\om\backslash\cup_i B_i} \ro^2 |\nab \varphi- A|^2+\hal \io |h-\he|^2 .
\end{eqnarray*}
As previously, since for the energy-minimizers $-\np h= (iu, \nab_A u)$, and $|\ro^2 -a_\ep |\le \frac{C}{\lep} $ in $ \om\backslash\cup_i B_i$, we have
$$\int_{\om\backslash\cup_i B_i} \ro^2 |\nab \varphi- A|^2= 
\int_{\om\backslash\cup_i B_i} \frac{|\nab h|^2}{a_\ep} (1-o(1)).$$
Therefore, in view of Proposition \ref{propballs}, 
$$
J_\ep(u_\ep,A_\ep)  \ge  \pi \sum_i |d_i| b(p_i) \lep (1-o(1)) +\int_{\om\backslash\cup_i B_i} \frac{|\nab h|^2}{a_\ep} (1-o(1)) +\io |h-\he|^2, $$
and with assertion 2) of Lemma II.3,
$$
\frac{J_\ep(u_\ep,A_\ep)}{\he^2}  \ge  \hal \frac{\lep}{\he} \io b |\mu_\ep| +\frac{1}{\he^2} \io \frac{|\nab \hb|^2}{a_\ep} +\io \left|\frac{\hb}{\he}-1\right|^2 -o(1).$$
We thus obtain, using assertion 3) of Lemma II.3 that 
\begin{equation}
\label{lbb}
\liminf \frac{J_\ep(u_\ep,A_\ep)}{\he^2} \ge \liminf \hal \( \frac{\lep}{\he} \io b |\mu_\ep|\) +\io \nab h_0\cdot \A\nab h_0 +|h_0-1|^2.\end{equation}
Similarly, using (\ref{lb1}), we obtain
\begin{equation}
\label{lba}
\liminf \frac{J_\ep(u_\ep,A_\ep)}{\he^2} \ge \liminf \hal \( \frac{\lep}{\he} \io a_\ep |\mu_\ep|\) +\io \nab h_0\cdot \A\nab h_0 +|h_0-1|^2.\end{equation}
Then, using the weak convergence of $\mu_\ep $ to $\mu_0$ in $\M$, and the weak lower semi-continuity of $\mu \mapsto \io b|\mu| $, we conclude from (\ref{lbb}) that
$$\liminf \frac{J_\ep(u_\ep,A_\ep)}{\he^2} \ge \frac{\Lambda}{2} \io b|\mu_0| +\io \nab h_0\cdot \A\nab h_0 +|h_0-1|^2= E(h_0). $$
\hfill $\Box$

\hfill

The final convergence result will then follow from the combination of this result with the upper bound of Section III, leading to the fact that necessarily $h_0$ has to  be $h_*$,  the minimizer of $E$, and $\mu_0=\mu_*$.

\section{Upper Bound}
In this section we prove Proposition~I.4. First we remark that if $h$ is the solution of $-\div(\mathcal A\nab h) + h = \mu$ with boundary value $1$, then  
$$h(x) - 1 = \int G(x,y) \,d(\mu-1)(y),$$
where $G(.,y)$ is the solution of $-\div(\mathcal A\nab h) + h = \d_y$ vanishing on $\bo$ and $\mu -1$ denotes the difference between the measure $\mu$ and the Lebesgue measure in $\Omega$. From this it follows easily that
\begin{equation}\label{eq}
\int_\om \nab h\cdot\mathcal A\nab h + |h-1|^2 = \iint G(x,y) \,d(\mu-1)(x)\,d(\mu-1)(y).
\end{equation}
This last expression will be the one we use.

To prove Proposition I.4 we will then need some properties of the Green functions $G_\ep$, $G_0$ associated to the operators $-\div(\Ae \nab u) + u$ and $-\div(\A \nab u) + u$ respectively. These properties will be proved at the end of this section.

\begin{lem} Let $a_\ep = b + \beta_\ep$ be a sequence of functions satisfying (H1) to (H4), and $\A$ be the homogenized limit of the matrices $\Ae = {a_\ep}^{-1} \Id$ as $\ep$ goes to zero. For any $y\in\om$, let $G_\ep(.,y)$ (resp. $G_0(.,y)$) be the solution of $-\div(\Ae \nab G_\ep) + G_\ep = \delta_y$ (resp. $-\div(\A \nab G_0) + G_0 = \delta_y$)  that vanishes on $\bo$.

The following properties hold:

1) $G_\ep(x,y)$, $G_0(x,y)$  are positive functions, and symmetric in $x$ and $y$.

2) $\Delta$ denoting the diagonal in $\mr^2$, there exists $C>0$ such that $G_\ep(x,y)$, $G_0(x,y)$ are bounded by
$$C\(\l|\log|x-y|\r| +1\)$$
for all $x,y\in \ol\om\times\ol\om\setminus\Delta$.

3) For any compact $K\subset \om$, there exists $C>0$ such that for any $x,y\in \om$ 
$$G_\ep(x,y) + \frac{a_\ep(x)}{2\pi} \log |x-y| \le \frac{C}{\eta(\ep)},$$
where $\eta(\ep)$ is defined in (H3).

4) $G_\ep$ converges to $G_0$ locally uniformly in $\ol\om\times\ol\om\setminus\Delta$.
\end{lem}
Then we have the following easy Lemma:
\begin{lem} 
The function
\begin{equation}\label{I}
I(\mu)=\frac{\La}{2}\int b \ d\mu +\hal\iint G_0(x,y)\,d(\mu-1)(x)\,d(\mu-1)(y)
\end{equation}
is sequentially lower semicontinuous over the set of positive Radon measures supported in $\ol\om$, with respect to weak-* convergence.
\end{lem}
The proof of this can be found in \cite{wermer} for instance. Note that $I(.)$ is well defined over the set of positive Radon measures if we admit the value $+\infty$. Note also that if we restrict to measures in $H^{-1}(\om)$ then \eqref{eq} shows that $I(\mu)$ is a lower semicontinous functional of $h = L^{-1} \mu$ where $L$ is the operator $u\to -\div(\mathcal A\nab u) + u$ defined on $H^1_0(\om)$. It follows that $I$ is a lower semicontinuous function of $\mu$ with respect to  $H^{-1}$ convergence.

Now the proof of Proposition I.4 splits into two propositions. 

\begin{pro} Assume that $\La >0$ and that (H1) to (H4) are satisfied.  Let $\mu$ be a positive  Radon measure with support in $\ol\om$ and $(p^i_\ep)_{1\le i\le n_\ep}$ be families of points in $\om$ such that $\forall i\neq j$
\begin{equation}\label{1}  |p^i_\ep - p^j_\ep| > 4 \ep, \quad d(\pie,\bo)>\alpha_0>0,
\end{equation}
where $\alpha_0$ is independent of $\ep$,
\begin{equation}\label{2}  \frac{2\pi}{\he} \sum_{i=1}^{n_\ep} \delta_{p^i_\ep}\quad \longrightarrow \quad \mu, \quad \text{ in the sense of measures}
\end{equation}
and 
\begin{equation}\label{3}\lim_{\ep\to 0}\( \sum_{\substack{i\neq j\\ |p^i_\ep - p^j_\ep| < \alpha}} \frac{\l|\log |p^i_\ep - p^j_\ep|\r|}{\he^2}\) \quad\xrightarrow[\alpha\to 0]{}\quad 0.
\end{equation}
Then there exist configurations $(v_\ep,B_\ep)_{\ep >0}$ such that
\begin{equation}\label{4}\limsup_{\ep\to 0}\frac{J_\ep(v_\ep,B_\ep)}{\he^2} \le \frac{\La}{2} \limsup_{\ep\to 0}\frac{\displaystyle 2\pi \sum_{i=1}^{n_\ep} a_\ep(p^i_\ep)}{\he} + \hal\iint G_0\ d(\mu-1)d(\mu-1),
\end{equation}
where $G_0$ is defined in Lemma III.1.
\end{pro}

This proposition states that under reasonable hypotheses on points $\pie$,  one can construct a good test configuration with prescribed vortices at $\pie$. Moreover, \eqref 2 implies that $n_\ep /\he$ is bounded. The following Proposition asserts that the construction of points $\pie$ is possible.

\begin{pro} 
Assume that $\La  >0$ and that (H1) to (H4) are satisfied. Then given any positive   Radon measure $\mu$ of the form $\sigma(x)\, dx$ where $\sigma$ is a positive continuous function compactly supported in $\om$, there exist families of points 
$(p^i_\ep)_{1\le i\le n_\ep}$ satisfying \eqref 1, \eqref 2, \eqref 3 and such that 
\begin{equation}\label{5}
\limsup_{\ep\to 0}\frac{\displaystyle 2\pi \sum_{i=1}^{n_\ep} a_\ep(p^i_\ep)}{\he} \le \io b(x)\, d\mu(x).
\end{equation}
\end{pro}

The proof of Proposition I.4 follows easily from these two Propositions. First, taking any positive Radon measure $\mu$ supported in $\ol\om$, we may approach it in the weak-* topology by measures $\mu_n = \sigma_n(x)\,dx$ where $\sigma_n\in C_c(\om)$ is a positive function. Applying  Propositions III.1 and III.2, we may construct test-configurations $(v_\ep^n,B_\ep^n)_{\ep>0}$ such that 

$$
\limsup_{\ep\to 0}\frac{J_\ep(v_\ep^n,B_\ep^n)}{\he^2} \le \frac{\La}{2} \int b(x)\, d\mu_n(x) + \hal\iint G_0\ d(\mu_n-1)d(\mu_n-1).
$$
Therefore the same inequality is satisfied if we replace $(v_\ep^n,B_\ep^n)$ by the minimizing configuration $(u_\ep,A_\ep)$. 
This proves that for each $n$,
$$
\limsup_{\ep\to 0}\frac{J_\ep(u_\ep,A_\ep)}{\he^2} \le I(\mu_n),
$$
and then, using Lemma III.2,
\begin{equation}\label{6}
\limsup_{\ep\to 0}\frac{J_\ep(u_\ep,A_\ep)}{\he^2} \le \frac{\La}{2}\io b \ d\mu +\hal\iint G_0(x,y)\,d(\mu-1)(x)\,d(\mu-1)(y).
\end{equation}
Using (\ref{eq}) we get the conclusion of Proposition I.4.

\subsection{Proof of Proposition III.1}

The method for constructing a test configuration $(v_\ep,B_\ep)$ with prescribed vortices $(p^i_\ep)_{1\le i\le n_\ep}$ follows closely that of \cite{ss3}. First we define $h_\ep$ to be the solution of
\begin{equation}\label{8}
\left\{
\begin{aligned}
  -\div(\Ae\nab h_\ep) + h_\ep &= \sum_{i=1}^{n_\ep} \mie &\quad & \mathrm{in}\ \om \\
   h_\ep &=\he &\quad & \mathrm{on} \ \bo,
\end{aligned}
\right.  
\end{equation}
where $\mie$ is the line element on the circle $\p B(\pie,\ep)$ normalized so that $\mie(\p B(\pie,\ep)) = 2\pi$. 

Then we let $B_\ep$ be any vector field such that $\curl B_\ep = h_\ep$. Finally, we define $v_\ep = \rho_\ep e^{i\varphi_\ep}$ as follows: first we let
\begin{equation}\label{6b}
\re(x) = \left\{  \begin{aligned}
         0  & \quad \text{if $|x-p_i^\ep|\le \ep$ for some $i$,} \\
         \sqrt{a_\ep(x)} \D \frac{|x-p_i^\ep|-\ep}{\ep} & \quad \text{if $\ep <|x-a_i^\ep| < 2\ep$ for some $i$,}\\
         \sqrt{a_\ep(x)} &\quad\text{otherwise,}
                 \end{aligned}
\right.  
\end{equation}
and for any $x\in\om_\ep=\om\setminus\cup_i B(p_i^\ep,\ep)$,
\begin{equation}
\label{7}
\pe(x) = \oint_{(x_0,x)} (B_\ep - \Ae\nabla^\perp h_\ep).\tau\,d\ell, 
\end{equation}
where $x_0$ is a base point  in $\om_\ep$, $(x_0,x)$ is any curve joining $x_0$ to $x$ in $\om_\ep$  and  $\tau$ is the tangent vector to the curve. From \eqref 8, we see that this definition of $\pe(x)$ does not depend modulo $2\pi$ on the particular curve $(x_0,x)$ chosen. The fact that $\pe$ is not defined on $\cup_i B(p_i^\ep,\ep)$ is not important since $\re$ is zero there. Thus, $\pe$ satisfies 
\begin{equation}
\label{9}
-\Ae\nabla^\perp h_\ep = \nabla\pe - B_\ep
\end{equation}
in $\om_\ep$. Having defined $v_\ep=\re e^{i\pe}$, we estimate $J_\ep(v_\ep,B_\ep)$. Recall that
\begin{equation}
\label{10}
J_\ep(v_\ep,B_\ep) = \hal\int_\om |\nab\re|^2 + \re^2|\nab\pe - B_\ep|^2 + |h_\ep - \he|^2 + \frac{1}{2\ep^2} \(a_\ep-\pe^2\)^2.
\end{equation}
Using the fact that $|\nab a_\ep|\ll \he$ (hypothesis (H2)) and that the number of points $\pie$ is less than $C\he$ --- which follows from \eqref 2 --- it is not difficult to check that 
\begin{equation}
\label{11}
\hal\io |\nabla\re|^2 + \frac{1}{2\ep^2} \(a_\ep-\re^2\)^2 \ll \he^2.
\end{equation}
Also,  from \eqref{6b}, \eqref 9, 
$$\re^2 |\nabla\pe - B_\ep|^2\le a_\ep|\nabla\pe - B_\ep|^2 =  \nabla h_\ep\cdot \Aep \nab h_\ep$$
in $\om_\ep$. Therefore, replacing in \eqref{10} and in view of \eqref{11}
\begin{equation}
\label{12}
\limsup_{\ep\to 0}\frac{J_\ep(v_\ep,B_\ep)}{\he^2}  \le \limsup_{\ep\to 0}\frac{1}{2\he^2} \io  \nabla h_\ep\cdot \Aep\nab h_\ep + |h_\ep - \he|^2.
\end{equation}

Because $h_\ep$ is the solution of \eqref 8, we may rewrite the right-hand side of this inequality as
\begin{equation*}
\limsup_{\ep\to 0}\hal \iint G_\ep(x,y)  \,d(\mue-1)(x)\,d(\mue-1)(y),
\end{equation*}
where 
\begin{equation}\label{13}
\mue = \frac{1}{\he} \sum_{i=1}^{n_\ep}\mie,
\end{equation}
and $\mie$ is defined in \eqref 8. It follows from \eqref 2, \eqref 8 and \eqref{13} that $\mue\to\mu$ as $\ep\to 0$. Thus, to finish the proof of the proposition, it remains to show that 
\begin{multline}\label{a}
\limsup_{\ep\to 0}\hal \iint G_\ep \,d(\mue-1)\,d(\mue-1)\le \frac{\La}{2} \limsup_{\ep\to 0}\frac{\displaystyle 2\pi \sum_{i=1}^{n_\ep} a_\ep(p^i_\ep)}{\he} \\ + \hal\iint G_0\ d(\mu-1)d(\mu-1)
\end{multline}

\subsubsection*{Proof of \eqref a}

Let $\alpha>0$ and let $\Delta_\alpha = \{(x,y)\mid |x-y|<\alpha\}$.  Recall that $\mue\to\mu$. Hence, it follows that $(\mue-1)\otimes(\mue-1)\to (\mu-1)\otimes(\mu-1)$ as $\ep\to 0$. But from Lemma~II.1, $G_\ep$ tends to $G_0$ uniformly in $\ol\om\times\ol\om\setminus\Delta_\alpha$, therefore
\begin{equation}\label{b}
\lim_{\ep\to 0}\hal \iint_{\ol\om\times\ol\om\setminus\Delta_\alpha} G_\ep  \,d(\mue-1)\,d(\mue-1) =  \hal \iint_{\ol\om\times\ol\om\setminus\Delta_\alpha} G_0  \,d(\mu-1)\,d(\mu-1).
\end{equation}
Now we treat the integral on $\Delta_\alpha$. More precisely we prove that
\begin{equation}\label{d}
\limsup_{\ep\to 0}\iint_{\Delta_\alpha} G_\ep \,d(\mue-1)\,d(\mue-1) \le \frac{\La}{2} \limsup_{\ep\to 0}\frac{\displaystyle 2\pi \sum_{i=1}^{n_\ep} a_\ep(p^i_\ep)}{\he} + o_\alpha(1),
\end{equation}
where $\lim_{\alpha\to 0}o_\alpha(1)=0$. Adding  \eqref b, \eqref d and letting $\alpha\to 0$ yields \eqref a.
We are left with proving \eqref d.
First we use the bound $|G_\ep(x,y)| < C \l|\log|x-y|\r|$ from which one easily gets
$$\iint_{\Delta_\alpha} G_\ep \,d(\mue-1)\,d(\mue-1) \le  \iint_{\Delta_\alpha} G_\ep \,d\mue\,d\mue + C \alpha^2 |\log\alpha|. $$
Therefore \eqref d will follow if we prove 
\begin{equation}\label{e}
\limsup_{\ep\to 0}\iint_{\Delta_\alpha} G_\ep \,d\mue\,d\mue \le \frac{\La}{2} \limsup_{\ep\to 0}\frac{\displaystyle 2\pi \sum_{i=1}^{n_\ep} a_\ep(p^i_\ep)}{\he} + o_\alpha(1).
\end{equation}
To prove this, we come back to the definition of $\mue$. From this definition, we have 
\begin{equation}\label{f}
\iint_{\Delta_\alpha} G_\ep \,d\mue\,d\mue \le \frac{1}{\he^2}\( \sum_{\substack{1\le i\neq j\le n_\ep\\ |p^i_\ep - p^j_\ep| < 2\alpha}} \iint G_\ep \,d\mie\,d\mje + \sum_{i=1}^{n_\ep} \iint G_\ep \,d\mie\,d\mie\).
\end{equation}
Let us first estimate the first sum on the right-hand side. If $x\in \text{Supp}\ \mie = \p B(\pie,\ep)$, $y\in\text{Supp}\ \mje$ and $i\neq j$, since $|p^i_\ep - p^j_\ep|>4\ep$, then $|x-y| > \hal |p^i_\ep - p^j_\ep|$. Using the bound $|G_\ep(x,y)| < C \l|\log|x-y|\r|$ together with the fact that 
 $|p^i_\ep - p^j_\ep|<2\alpha$ and $\alpha$ is small enough, we get
$$\iint G_\ep \,d\mie\,d\mje <  C \l|\log |p^i_\ep - p^j_\ep|\r|. $$
Then, by hypothesis \eqref 3,
\begin{equation}\label{g}
\limsup_{\ep\to 0}\frac{1}{\he^2} \sum_{\substack{1\le i\neq j\le n_\ep\\ |p^i_\ep - p^j_\ep| < 2\alpha}} \iint G_\ep \,d\mie\,d\mje \le o_\alpha(1).
\end{equation}
As for the second sum in the right-hand side of \eqref f, we use property 3) in Lemma~III.1 to get that for any $1\le i\le n_\ep$, and any $x,y\in\text{Supp}\ \mie$, 
\begin{equation}\label{h}
G_\ep(x_,y) + \frac{a_\ep(x)}{2\pi} \log |x-y|  < \frac{C}{\eta(\ep)}\ll \lep.
\end{equation}
But $x\in\text{Supp}\mie$ is equivalent to $|x-\pie|=\ep$. Then property (H2) of $a_\ep$ implies that $a_\ep(x) \approx a_\ep(\pie)$ as $\ep\to 0$. Replacing in \eqref h and integrating w.r.t. $\mie\otimes\mie$ yields
$$\iint G_\ep \,d\mie\,d\mie \le 2\pi a_\ep(\pie)\lep \(1+o_\ep(1)\)$$
and then, summing over $1\le i\le n_\ep$ and dividing by $\he$,
\begin{equation}\label{i}
\limsup_{\ep\to 0}\frac{1}{\he^2}\sum_{i=1}^{n_\ep} \iint G_\ep \,d\mie\,d\mie \le \frac{\La}{2} \limsup_{\ep\to 0}\frac{\displaystyle 2\pi \sum_{i=1}^{n_\ep} a_\ep(p^i_\ep)}{\he}.
\end{equation}
Here we have used the fact that $\lep\thicksim \La\he$. 
Thus \eqref e is proved and the Proposition follows.
\hfill $\Box$

\subsection{Proof of Proposition III.2}
Let $\mu = \sigma(x)\, dx$, $C = \|u\|_{\infty}$ and $\alpha_0 = dist(\text{supp}\mu,\bo)$.  Also, let 
\begin{equation}\label{ot}
\widetilde\om=\{x\in\om\mid d(x,\bo)> \alpha_0/2\}.
\end{equation}
Recall that from hypothesis (H3) on $a_\ep$ there exists a positive function $\delta(\ep)$ such that 
\begin{equation}\label{j}
\d(\ep)\ll \frac{1}{(\llep )^\hal},\quad \text{and for any $x\in\om$, $\min_{B(x,\d(\ep))} \beta_\ep = 0$}.
\end{equation}

For any $\ep>0$, we tile $\mr^2$ with open squares of sidelength $2\delta(\ep)$ and let $\K(\ep)$ be the family of those squares that are entirely inside $\widetilde\om$. We denote by $c_K$ the center of a square $K$. Since $\mu$ is absolutely continuous with respect to the Lebesgue measure, we have $\mu (K)\leq C \delta^2$.

Now the family of points $(\pie)_{1\le i\le n_\ep}$ is defined as follows: for any $K\in \K(\ep)$, we let
\begin{equation}\label{k}
n(K,\ep) = \left[\frac{\he(\ep)\mu(K)}{2\pi}\right],
\end{equation}
where $[x]$ is the biggest integer  no greater than $x$. Using \eqref j there is a point $p_K\in B(c_K,\d)$ such that $\beta_\ep(p_K) = 0$ ($p_K$ is a pinning site). We now pick $n(K,\ep)$ points evenly scattered in the ball $B(p_K,1/\he)$, and  we call $\P(K,\ep)$ their union. By evenly scattered we mean that for any $p,q\in\P(K,\ep)$,
\begin{equation}\label{l}
|p-q|\ge \frac{C}{\he\sqrt{n(K,\ep)}}.
\end{equation}
We let 
\begin{equation}\label{m}
n_\ep = \sum_{K\in\K(\ep)} n(K,\ep),\quad \text{and } \P(\ep) = \cup_{K\in\K(\ep)} \P(K,\ep) = (\pie)_{1\le i\le n_\ep}
\end{equation}
be our family of points. We now check that this family satisfies \eqref 1, \eqref 2, \eqref 3 and \eqref 5. 

\eqref 1 is clear from \eqref l if $\pie, \pje$ belong to the same pinning site. It is even more true if $\pie, \pje$ do not belong to the same site since in this case their mutual distance is at least $2\d(\ep) \gg \ep$. Moreover  from \eqref{ot} we have $d(\pie,\bo)>\alpha_0/2$.

For \eqref 2, let 
\begin{equation}\label{n}
\mue = \frac{2\pi}{\he}\sum_{i=1}^{n_\ep} \d_{\pie}
\end{equation}
and $f$ be a continuous function in $\ol\om$. We let 
$\gamma_\ep = \sup_{K\in\K(\ep)}\sup_{x,y\in K} |f(x) - f(y)|$. Then since the size of the squares in $\K(\ep)$ tends to zero with $\ep$, so does $\gamma_\ep$. Let $K_\ep$ be the union of the squares in $\K(\ep)$, then for $\ep$ small enough $\text{supp}\mu\subset K_\ep$ and 
$$\left|\int f\,d\mu - \int f\,d\mue\right|\le 
\|f\|_\infty \sum_{K\in\K(\ep)}|\mu(K)-\mue(K)| + \gamma_\ep (\mue+\mu)(K_\ep).
$$
It is clear that the second  term on the right-hand side goes to zero with $\ep$. For the first term we note that from \eqref k, \eqref n, we have $|\mu(K)-\mue(K)|\le 2\pi/\he$ while the number of squares in $\K(\ep)$ is of the order of $1/\d^2$. From \eqref j it then follows that $\sum_{K\in\K(\ep)}|\mu(K)-\mue(K)|$ tends to zero with $\ep$. We thus have $\lim_{\ep\to 0}\int f\,d\mue=\int f\,d\mu $ and   \eqref 2 follows.

 We easily deduce  \eqref 5 from \eqref 2. Indeed from (H2) and the fact that each point is at a distance  at most $1/\he$ from a pinning site, we get that $a_\ep(p)\approx b(p)$ as $\ep\to 0$, uniformly in $p\in\P(\ep)$. Moreover, since $n_\ep / \he$ is bounded,
$$\lim_{\ep\to 0}\frac{\displaystyle 2\pi \sum_{i=1}^{n_\ep} a_\ep(p^i_\ep)}{\he}  =  \lim_{\ep\to 0}\frac{\displaystyle 2\pi \sum_{i=1}^{n_\ep} b(p^i_\ep)}{\he} = \int b(x)\, d\mu(x),$$
by the convergence of $\mue$ to $\mu$.

It remains to prove \eqref 3. We split the sum in \eqref 3 as follows: let $\I(\ep)$ be the set of pairs of indices $(i,j)$ such that $1\le i\neq j\le n_\ep$ and $\pie,\pje$ belong to the same square of the subdivision $\K(\ep)$. Let $\J(\ep)$ be pairs $(i,j)$ such that $\pie,\pje$ belong to different squares. Then
\begin{equation}\label{o}
\sum_{\substack{i\neq j\\ |p^i_\ep - p^j_\ep| < \alpha}} \l|\log|p^i_\ep - p^j_\ep|\r| = 
\sum_{\substack{(i,j)\in \I(\ep)\\|p^i_\ep - p^j_\ep| < \alpha}} \l|\log |p^i_\ep - p^j_\ep|\r| +
\sum_{\substack{(i,j)\in\J(\ep)\\ |p^i_\ep - p^j_\ep| < \alpha}} \l|\log |p^i_\ep - p^j_\ep|\r|
\end{equation}
The first sum  in \eqref o is estimated as follows. For every $K\in\K(\ep)$, $\mu(K) < C \d^2$ thus the number of points of $\P(\ep)$ in $K$ is less than $C\d^2\he$. The number of squares  being of the order of $\d^{-2}$, the cardinal of $\I(\ep)$ is less than $C\d^2\he^2$. Using \eqref j, \eqref k and \eqref l, we find
\begin{equation}\label{p}
\sum_{\substack{(i,j)\in \I(\ep)\\|p^i_\ep - p^j_\ep| < \alpha}} \l|\log |p^i_\ep - p^j_\ep|\r|\le C \he^2\d^2\llep \ll \he^2.
\end{equation}

To treat the second sum in \eqref o, we note that if $K$ and $K'$ are distinct squares in $\K(\ep)$ and $p\in K$, $q\in K'$ then 
$$\forall x\in K, \forall y\in K',\quad |x-y|\le 4|p-q|. $$
Thus we may write, using the fact that $\mu(K)<C\d^2$,
$$
\sum_{\substack{i\neq j\\ p^i_\ep\in K,  p^j_\ep\in K'}} \l|\log |p^i_\ep - p^j_\ep|\r| \le C\he^2 \iint_{K\times K'}(\l|\log|x-y|\r| +1 )\,dx\,dy.
$$
Summing over pairs of squares $K,K'\in\K(\ep)$ such that $K\times K'$ intersects $\{(x,y)\mid |x-y|<\alpha\}$ we get for $\ep$ small enough
\begin{equation}\label{q}
\sum_{\substack{(i,j)\in\J(\ep)\\ |p^i_\ep - p^j_\ep| < \alpha}} \l|\log |p^i_\ep - p^j_\ep|\r|\le  C \he^2\iint_{|x-y|< 2\alpha} (\l|\log|x-y|\r| +1) \,dx\,dy.
\end{equation}
Summing \eqref p, \eqref q, dividing by $\he^2$ and letting $\ep$ and then $\alpha$ tend to zero yields \eqref 3. Proposition III.2 is proved.
\hfill $\Box$

\subsection{Proof of Lemma III.1}

The fact that $G_\ep$ and $G_0$ are positive is a simple consequence of the maximum principle, that they are symmetric is standard and follows from Green's identity.

The inequality 
$$ G_\ep(x,y), G_0(x,y) < - C \log|x-y| + C$$
is a well known property of Green functions for elliptic operators  in divergence form, a proof can be found in \cite{st}.

To prove property 3), we let 
$$v_\ep(x,y) = G_\ep(x,y) +\frac{a_\ep(y)}{2\pi}\log|x-y|$$
and $L_\ep$ be the operator $u\mapsto -\div(\Ae\nabla u) + u$. Then letting $f_\ep = L_\ep v_\ep (.,y)$, we have 
\begin{equation}\label{r}
f_\ep (x,y) = -\frac{a_\ep(y)}{2\pi} \nabla\frac{1}{a_\ep(x)}.\nabla_x \log|x-y| - \frac{a_\ep(y)}{2\pi} \log|x-y|.
\end{equation}
Thus for any $1\le q<2$, there is a $C$  independent of $y$ and $\ep$, such that $\|f_\ep(.,y)\|_{L^q}\le C/\eta(\ep)$. On the other hand, $v_\ep(.,y)$ is bounded in $W^{1,q}(\om)$ independently of $\ep$ and $y$ (see \cite{st}). 

Now, Theorem 2 of \cite{me} implies that there exist $p>2$ and $p'<2$ such that if $u$ satisfies $L_\ep u = f$, then for any compact $K\subset \om$,
$$\|\nabla u\|_{L^p(K)} \le C(K) \(\|\nabla u\|_{L^{p'}(\om)} + \|f\|_{W^{-1,p}(\om)}\).$$
We may choose $q<2$  such that $W^{-1,p}\subset L^q$ and $p'<q$. Thus, we find that $v_\ep(.,y)$ is bounded in 
$W^{1,p}(K)$ by $C/\eta(\ep)$. Since $p>2$, this yields the uniform bound $\forall x\in K, \forall y\in\om$, 
$$|v_\ep(x,y)|\le \frac{C(K)}{\eta(\ep)}$$
i.e. property 3).

To prove property 4), we note that for any $\alpha>0$,  $L_\ep G_\ep(.,y) = 0$ in $\om\setminus B(y,\alpha)$ while $G_\ep(.,y)$ is bounded in $W^{1,q}(\om)$ independently of $\ep$ and $y$ (see \cite{st}). Using the aforementioned result of \cite{me}, we find that $G_\ep(.,y)$ is bounded in $W^{1,p}_{\loc} (\om\setminus B(y,\alpha))$, for some $p>2$, independently of $y$ and $\ep$, thus $G_\ep$ converges locally uniformly in $\om\times\om\setminus \Delta$, where $\Delta$ is the diagonal. The limit is necessarily $G_0$, since  $G_0(.,y)$ satisfies $L_0 G_0(.,y) = -\div \A \nabla_x G_0 + G_0 = \delta_y$ and  $L_\ep$ $H$-converges to $L_0$. Lemma II.1 is proved.
\hfill $\Box$


\section{Convergence results}

We can then proceed as in the rest of Section III in \cite{ss3}. 
\begin{pro}
The minimum of $E$ is uniquely achieved by $h_*\in C^{1, \gamma}(\om) (\forall \gamma<1)$  satisfying 
\begin{equation}
\label{I32}
\left\{
\begin{array}{ll}
h_* \ge 1-\D\frac{\Lambda b}{2} & \mathrm{in }\  \om\\
h_*=1 & \mathrm{on }\ \bo\\
\mu_*:= -\mathrm{div} (\A \nab h_*) +h_* \ge 0 \\
\( h_* -\(1-\D\frac{\Lambda b}{2} \)\) \mu_* =0
\end{array}
\right. \end{equation}
\end{pro}
As in \cite{ss3}, we divide the proof of this proposition into several lemmas. 
\begin{lem}
Let $\mu_*^+ $ and $\mu_*^-$ be the positive and negative parts of the measure $\mu_*$. Then \begin{eqnarray*}
h_* = 1-\frac{\Lambda b}{2} & \quad \mu_*^+ \text{ a.e.}\\
h_* = 1+\frac{\Lambda b}{2} & \quad \mu_*^- \text{ a.e.}
\end{eqnarray*}
$$ 1-\frac{\Lambda b}{2} \le h_* \le 1+\frac{\Lambda b}{2} .$$
\end{lem}
{ \it Proof :} As in \cite{ss3}, the minimum of $E$ is achieved by some $h_*$, by lower semi-continuity. Performing variations $(1+tf)\mu_*$ where $f\in C^0(\om)$, and looking at the first order in $t\to 0$, we find similarly as in \cite{ss3} that
$$\frac{\Lambda b}{2} |\mu_*| +(h_*-1) \mu_* =0 .$$
Hence, 
\begin{eqnarray*}
h_* = 1-\frac{\Lambda b}{2} & \quad \mu_*^+ \text{ a.e.}\\
h_* = 1+\frac{\Lambda b}{2} & \quad \mu_*^- \text{ a.e.}.
\end{eqnarray*}
As in \cite{ss3}, considering variations $\mu_* +\nu$, where $\nu\in \M\cap H^{-1}$ and $\nu$ and $\mu_*$ are mutually singular, we are led to $ 1-\frac{\Lambda b}{2} \le h_* \le 1+\frac{\Lambda b}{2}.$
\hfill $\Box$

\begin{lem}
$\mu_*$ is a positive measure.
\end{lem}
{\it Proof :} 
$$\io \mu_* (h_*-1)_+=  \io\mu_*^+(h_*-1)_+ - \io \mu_*^-(h_*-1)_+ .$$
Since   $(h_*-1)_+=0$ $\mu_*^+$-a.e., we have 
\begin{eqnarray*}
\io \mu_* (h_*-1)_+ & = & - \io \mu_*^-(h_*-1)_+\\
& = & \io (-\div (\A \nab h_*) +h_*)(h_*-1)_+\\
& = & \int_{h_* >1} \nab h_*\cdot( \A \nab h_*) + h_*(h_*-1) \ge 0,\end{eqnarray*}
because $\A$ is a symmetric positive matrix (this follows from the compactness of the set of matrices bounded from above and below). 
We deduce that $$\io \mu_*^-(h_*-1)_+ =0,$$
but since $h_*-1= \frac{\Lambda b}{2}$, $\mu_*^-$ a.e., we have
$$\io 
\frac{\Lambda b}{2} \mu_*^- =0,$$
hence $\mu_*^-=0$, and $\mu_* \ge 0$.
\hfill $\Box$

\hfill

Thus, $h_*$ satisfies all the properties listed in (\ref{I32}).\\

 We can now complete the convergence results. From the upper bound of Proposition I.4 and Lemma II.4, we deduce that for our family of minimizers $(u_\ep, A_\ep)$,
$$
\min_V E = E(h_*) \ge \liminf_{\ep \to 0} \frac{J_\ep(u_\ep,A_\ep)}{\he^2} \ge E(h_0) \ge E(h_*).$$
$h_*$ being the unique minimizer of $E$, we conclude that $h_0= h_*$ and thus $\mu_0 = \mu_*$. We also obtain 
\begin{equation}
\label{cv}
\lim_{\ep \to 0} \frac{J_\ep(u_\ep,A_\ep)}{\he^2}=E(h_*).\end{equation}
Since the possible limits are unique, the whole family $\frac{h_\ep}{\he}$ converges to $h_*$, and the same for $\mu_\ep$. 

 In view of (\ref{lba}), we have
\begin{eqnarray*}
\liminf_{\ep \to 0} \frac{J(u_\ep, A_\ep)}{\he^2} & \ge & 
\liminf_{\ep \to 0} \hal \( \frac{\lep}{\he} \io a_\ep |\mu_\ep|\) +\hal\io \nab h_*\cdot \A\nab h_* +|h_*-1|^2.\\
& \ge & \frac{\Lambda}{2}\io b|\mu_*|+\hal\io \nab h_*\cdot \A\nab h_* +|h_*-1|^2,\end{eqnarray*}
while 
$$\limsup_{\ep \to 0}\frac{J(u_\ep, A_\ep)}{\he^2} \le \frac{\Lambda}{2}\io b|\mu_*|+\hal\io \nab h_*\cdot \A\nab h_* +|h_*-1|^2.$$
Thus, we deduce that 
$$\lim_{\ep \to 0} \io a_\ep |\mu_\ep|= \io b\mu_*.$$
On the other hand, $$\liminf_{\ep \to 0}\io a_\ep |\mu_\ep|\ge \liminf_{\ep \to 0} \io b |\mu_\ep |\ge \io b|\mu_*|,$$
hence $ \io b|\mu_\ep |\to \io b \mu_*$, while $\io b\mu_\ep \to \io b\mu_*$. 
We conclude that $\io b(|\mu_\ep|- \mu_\ep)\to 0$ and thus $ |\mu_\ep|$ and $\mu_\ep$ have the same limiting measure $\mu_*$. This proves 
(\ref{convaepi}), (\ref{ddelta}), and (\ref{|d|delta}).
 
\noi
Following \cite{ss3}, Section IV, we can also prove easily the following : 
\begin{pro} 
If $\Lambda =0$, then $h_*=1$ and $\frac{h_\ep}{\he }-1 \to 0$ strongly in $H^1_0(\om)$. If $\Lambda>0$, then $\frac{h_\ep}{\he}-1\rightharpoonup h_*-1 $ in $H^1_0(\om)$, the convergence is not strong  and 
$$\frac{|\nab h_\ep|^2 }{\he^2 a_\ep} \to \nab h_* \cdot \A \nab h_* + \Lambda b \mu_* \quad \text{in } \M.$$
\end{pro}
{\it Proof :} First, it is easy to get, as seen in Lemma II.4 for example, that
$$ \io |\nab_{A_\ep} u_\ep|^2 \ge \io \frac{|\nab h_\ep|^2}{a_\ep } (1-o(1)), $$
thus, we have 
\begin{eqnarray}
\liminf_{\ep \to 0} \frac{J(u_\ep, A_\ep)}{\he^2} & \ge & \liminf_{\ep \to 0}\frac{1}{\he^2}\left(\hal \io  \frac{|\nab h_\ep|^2}{a_\ep }+|h_\ep-\he|^2\right)\\
\label{nha}
& \ge & 
 \frac{\Lambda}{2} \io b \mu_* + \hal \io \nab h_* \cdot \A \nab h_*+|h_*-1|^2.
 \end{eqnarray}
The case $\Lambda =0$ follows easily from the upper bound $\min J_\ep(u_\ep,A_\ep)\le o(\he^2)$  of Section II combined with (\ref{nha}).

The convergence of $\frac{h_\ep}{\he} $ to $h_*$ is weak in $H^1$, in general, thus strong in $L^2(\om)$, and 
$$\lim_{\ep \to 0}\io \left|\frac{h_\ep}{\he}-1\right|^2 = \io |h_*-1|^2.$$
Combining this to the convergence result (\ref{cv}), we have
\begin{equation}
\label{fin}
\lim_{\ep \to 0} \hal \io \frac{|\nab h_\ep|^2}{\he^2 a_\ep} = \frac{\Lambda}{2} \io b \mu_* + \hal \io \nab h_* \cdot \A \nab h_*.
\end{equation}
Then, we argue as in \cite{ss3}, Proposition IV.1.
Roughly speaking, one considers any open set $U\subset \om$, and gets a lower bound 
\begin{eqnarray*}
\liminf_{\ep\to 0} \int_U \frac{|\nab h_\ep|^2}{\he^2 a_\ep} & = & \liminf_{\ep\to 0}\int_{U \cap(\cup_i B_i)} \frac{|\nab h_\ep|^2}{\he^2 a_\ep}+ \int_{U \backslash \cup_i B_i} \frac{|\nab h_\ep|^2}{\he^2 a_\ep}\\
& \ge & \Lambda \int_U b|\mu_\ep|+ \int_U \nab h_* \cdot \A \nab h_*\\
& \ge & \Lambda \int_U b\mu_*+  \int_U \nab h_* \cdot \A \nab h_*. \end{eqnarray*}
Since this is true for any $U \subset \om$, comparing this to (\ref{nha}) and (\ref{fin}), we obtain as in \cite{ss3},
$$\frac{|\nab h_\ep|^2 }{\he^2 a_\ep} \to \nab h_* \cdot \A \nab h_* + \Lambda b \mu_* \quad \text{in } \M.$$
\hfill $\Box$\\
\noindent
This completes the proof of Theorems 1, 2 and 3.


\hfill

\noindent
{\it Acknowledgments :} The authors are very grateful to Fran\c cois Murat for taking time explaining the basis of homogenization and pointing out the good references. They would also like to thank very much Jon Chapman for fruitful discussions on pinning models and Alano Ancona for pointing out references on Green functions.

\end{document}